**A joint longitudinal-survival framework for dynamic treatment regimen evaluation in sequential multiple assignment randomized trials**

Zhengxi Chen M.S.[1] and Holly Hartman Ph.D.[1]

[1]Department of Population and Quantitative Health Sciences, Case Western Reserve University School of Medicine, Cleveland, OH, USA

**Correspondence**: Zhengxi Chen M.S.

Department of Population and Quantitative Health Sciences

Case Western Reserve University School of Medicine

10900 Euclid Avenue, Cleveland, OH 44106

Email: zxc693@case.edu

Tel: +1 608-698-7918

Fax: +1 216-368-4547

**Abbreviated title:** Joint longitudinal-survival framework for DTR evaluation

**Abstract**

Sequential multiple assignment randomized trials (SMARTs) provide a systematic framework for constructing and evaluating dynamic treatment regimens (DTRs). In clinical studies, longitudinal biomarkers are routinely collected to monitor disease progression and define treatment response. However, the integration of longitudinal biomarker data into survival analysis for DTR evaluation within a SMART remains unexplored. We propose a joint longitudinal-survival framework to estimate DTR-specific survival outcomes within a two-stage SMART. A linear mixed model specifies the biomarker trajectory, and a relative risk model links the survival process to the current latent biomarker value. To accommodate the time-varying treatment assignment, treatment effects are parameterized through piecewise cumulative exposure terms with a structural change at the decision point. Joint-model parameters are estimated by maximum likelihood using pseudo-adaptive Gauss–Hermite quadrature for random-effect integration. Under standard causal identification assumptions, DTR-specific marginal survival outcomes are obtained through a plug-in parametric g-formula. We compare the joint-model framework to the inverse probability of treatment weighted Kaplan–Meier estimator and implement the multiple comparisons with the best method to identify the optimal DTR with multiplicity control. Through a simulation study and an application to a SMART for androgen-independent prostate cancer, we demonstrate that the joint-model framework produces unbiased estimates under correct model specification, with substantial efficiency gains and higher accuracy in optimal DTR identification. This framework exploits prognostic information from longitudinal biomarkers, enables valid causal inference for DTR-specific survival outcomes, and provides a generalizable analytic tool for developing evidence-based adaptive treatment strategies.

## 1.Introduction

Dynamic treatment regimens (DTRs) are sequential decision rules that formalize personalized treatment strategies based on efficacy outcomes from earlier treatments.[1] Specifically, they map patients' accumulated clinical and treatment history to adapt subsequent treatment at each stage of intervention, with decision points referring to the time points when treatment adaptations occur.[2,3] Decision rules are informed by tailoring variables that capture patients' evolving health status and guide the selection among available intervention options, which may include changes in treatment type, dosage, timing, combination, or continuation of the current therapy.[2,3] DTRs provide an effective strategy for the management of chronic diseases, where clinicians repeatedly evaluate patients' evolving status and make dynamic treatment decisions over an extended period.[4] DTRs have been constructed and applied in substance use disorders, diabetes, and HIV/AIDS.[1,5]

Improvements in risk factor identification, earlier diagnosis, and therapeutic advances have enabled many cancers to be managed as chronic diseases through stable long-term control, such as breast cancer with effective treatments to delay progression and prostate cancer that is naturally indolent with slow progression and stable localization.[6,7] In oncology studies, DTRs naturally arise when an initial therapy is followed by a response-dependent adaptation to maintenance, salvage, or dose modification.

Sequential multiple assignment randomized trials (SMARTs) provide an experimental framework for constructing and evaluating DTRs.[2,4,8] In SMARTs, participants are followed through multiple treatment stages and randomized to available treatment options at each stage based on their response to previous treatments. Each stage corresponds to a decision point in a DTR, and the collection of treatment sequences across stages defines a set of embedded decision rules. SMART designs have been applied across a wide range of hematologic and solid tumor malignancies, including acute myelogenous leukemia, lymphoma, multiple myeloma, lung cancer, prostate cancer, melanoma, and renal cell carcinoma.[9]

Overall survival is broadly recognized as the gold standard endpoint in oncology trials, as extending survival is the primary goal of cancer treatment.[10] Time-to-event outcomes appropriately accommodate situations where the event of interest is not observed in all participants through censoring due to study completion or loss to follow-up.[11] However, in multi-stage trials, the sequential nature of treatment assignment, treatment switching, and delayed effects can create informative-censoring patterns that bias conventional survival comparisons.[12-15]

Many studies have examined survival outcomes within DTRs and multi-stage randomized trial frameworks. The earliest studies implemented separate assessment of first- and second-stage treatments using the Kaplan–Meier estimator and the Cox Proportional Hazards model.[16-20] These fragmented approaches are inadequate to evaluate survival outcomes across different DTRs since they do not incorporate patients' full treatment histories or appropriately account for the randomization mechanisms at each stage.

To address these limitations, numerous estimators have been developed using inverse probability weighting (IPW) to adjust for multiple randomizations and compare survival outcomes across DTRs.[21] Lunceford et al. introduced estimators for marginal mean restricted survival time and survival distributions based on the marginal mean model proposed by Murphy et al.[22,23] To correct the bias inherent in subgroup-specific Kaplan–Meier estimators, Wahed and Tsiatis developed semiparametric weighted Kaplan–Meier estimators.[24,25] Subsequently, estimators were designed specifically to compare overall survival between DTRs that follow completely separate treatment sequences. Guo and Tsiatis developed a weighted risk set estimator to estimate cumulative hazard functions and tested the equality of survival curves using a weighted log-rank test.[26] However, it was later shown to have limited power for detecting time-varying hazard functions. Feng and Wahed extended this work with a supremum inverse-probability-of-randomization-weighted log-rank test to improve sensitivity.[27] To accommodate DTRs with shared first-stage treatments, Kidwell and Wahed proposed a univariate weighted log-rank test.[28] Additional contributions include maximum likelihood estimation through mixture distributions,[29] fixed and time-dependent weighted Kaplan–Meier estimators,[30] doubly robust estimators of survival outcomes,[31,32] and extension to competing risks through the cumulative incidence function.[33] A common limitation of these methods is that they focus on comparing averaged survival outcomes across DTRs without parameterizing treatment effects or incorporating auxiliary variables such as baseline patient characteristics.

To enable parameterization of treatment effects within DTRs, Lokhnygina and Helterbrand extended the Cox model using a weighted pseudo-score equation.[34] However, this approach is restricted to DTRs with fixed second-stage treatment and does not accommodate auxiliary variables. Subsequent approaches extended the survival model to incorporate auxiliary variables. For example, Tang and Wahed introduced a stratified model that assumes proportional hazards within each DTR relative to baseline covariates, with treatment comparisons based on the log ratios of estimated cumulative hazard functions.[35] Although this method accommodates non-proportional hazards across treatment regimens, it relies on restrictive proportional hazards assumptions that preclude time-varying effects and treatment-covariate interactions. Furthermore, Thall et al. presented a Bayesian framework that models the second failure time conditional on the intermediate worsening time with time-varying hazards.[36] However, this approach restricts the survival distribution to the Weibull family and is susceptible to overparameterization. More recently, Chao et al. proposed a conditional hazard-based joint model that simultaneously analyzes time to response and time to death, which enables estimation of baseline covariate effects, treatment effects, treatment-covariate interactions within DTRs, as well as time-dependent covariates measured at the initiation of second-stage treatment.[37]

Despite extensive methodological progress in survival estimation within DTRs, no existing method links longitudinal biomarker process to the survival outcome within a unified framework. In oncology research, longitudinal biomarkers, such as circulating tumor cells, immunologic indicators, and clinical biomarkers, serve as natural candidates to monitor disease progression, define response criteria for previous treatment, and guide subsequent treatment adaptation. For example, Bianchi et al. conducted a two-stage trial for metastatic castration-resistant prostate cancer, where prostate-specific antigen (PSA) was measured repeatedly and only patients with a PSA response (> 50% reduction from baseline) after four chemotherapy cycles were eligible for the second-stage treatment randomization.[16] Beyond their role in defining response, longitudinal biomarkers may also serve as time-varying confounders, since the biomarker trajectory is influenced by treatment sequence and simultaneously associated with the hazard of the event. Moreover, the dynamic nature of biomarkers, including their values and the rate of change over time, provides valuable prognostic information that extends beyond baseline measurements.

The conventional Cox model can incorporate time-dependent covariates through a counting process framework. However, it relies on several strong assumptions[38]: (1) covariates must be external to the survival process, meaning their future values are known in advance and unaffected by the event; (2) covariates are measured continuously over time without error; and (3) covariates change only at scheduled time points and remain constant between visits. These assumptions are not appropriate for longitudinal biomarkers in oncology studies, which are typically internal, prone to measurement error, collected intermittently, and may have delayed or cumulative effects on the event.

Joint models of longitudinal and time-to-event data are designed to address these challenges by linking a longitudinal process for the biomarker trajectory with a survival process through shared subject-specific latent structure.[38] In practice, the latent structure comprises random effects obtained from the longitudinal submodel that enter the survival submodel as predictors of the hazard. This unified approach offers several methodological advantages: (1) it accounts for informative censoring when biomarker measurements are terminated by the event since the shared random effects make the missing data mechanism ignorable within the joint likelihood; (2) it reduces estimation bias from measurement errors by modeling the true longitudinal trajectory; (3) it improves statistical efficiency by utilizing all available data within a single likelihood function; and (4) it facilitates comprehensive inference on treatment effects through both the longitudinal and survival processes.[39,40] Prior studies have demonstrated that, compared with the conventional Cox model, joint models provide more efficient and less biased estimates of single-stage treatment effects with reduced standard errors, indicating that studies using joint models require smaller sample sizes while maintaining proper statistical power.[41] However, the implementation of joint model within multi-stage trial designs for estimating regimen-specific causal effects remains unexplored.

In this study, we propose a joint longitudinal-survival framework to estimate regimen-specific survival outcomes under a two-stage tailoring variable-SMART design. In Section 2, we present the SMART design that motivates this study. Section 3 specifies the joint model for longitudinal and survival outcomes. In Section 4, we describe the joint likelihood construction and the maximum likelihood estimation procedures. Section 5 outlines the causal identification and inference procedures for estimating and comparing DTRs. Section 6 reports our simulation study evaluating the performance of the proposed framework. An application of the proposed framework to a SMART for androgen-independent prostate cancer is presented in Section 7, and we conclude with a discussion in Section 8.

**2. SMART design**

We consider a two-stage tailoring-variable SMART design with survival endpoints. For simplicity, each decision point offers two intervention options. However, SMARTs can be extended to include more than two stages, and three or more intervention options at any decision point.

At study entry, each patient is randomized with equal probability ($P_1 = 0.5$) to one of two first-stage treatments, $V_1 \in \{\text{A}, \text{B}\}$. For patient $i = 1, \ldots, N$, let $y_{ij}$ denote the longitudinal biomarker measured at time $t_{ij}$ for visit $j = 0, 1, 2, \ldots, n_i$, where $N$ is the total number of subjects and $n_i$ is the number of repeated measurements after baseline for patient $i$. At the pre-specified decision point $\tau$, patients who experience the survival event before $\tau$ are not eligible for second-stage treatment. Patients who survive to $\tau$ are evaluated for response based on the observed biomarker value change from baseline to $\tau$. We assume that higher values of the longitudinal biomarker correspond to worse clinical status. A patient is classified as a responder if the observed reduction in the longitudinal biomarker from baseline to time $\tau$ meets or exceeds a pre-specified threshold $c$, and the response indicator is denoted as $R_i = I(y_{i0} - y_{i\tau} \geq c)$. Responders continue their first-stage treatment ($V_2 = V_1$), while non-responders are randomized with equal probability ($P_2 = 0.5$) to one of two alternative second-stage treatments, $V_2 \in \{\text{C}, \text{D}\}$.

Each DTR $g$ is characterized by a three-part structure that outlines the sequence of recommended interventions in the order of first-stage treatment, second-stage treatment for responders, and second-stage treatment for non-responders, denoted as $g = (V_1, V_2(R = 1), V_2(R = 0))$. Our design embeds four DTRs: (A, A, C), (A, A, D), (B, B, C), and (B, B, D). Figure 1 illustrates the randomization structure of the SMART design, longitudinal biomarker measurement schedule, and survival follow-up through administrative censoring.

## 3. Joint longitudinal-survival model

### 3.1 Longitudinal submodel

The longitudinal process is modeled through a linear mixed model (LMM),

$$y_{ij} = m_i(t_{ij}) + \varepsilon_{ij} = x_i(t_{ij})^\top \beta + z_i(t_{ij})^\top b_i + \varepsilon_{ij},$$

where the latent mean trajectory $m_i(t_{ij})$ decomposes the biomarker evolution into fixed population-average and subject-specific components. Here, $x_i(t_{ij})$ denotes a vector of fixed-effect covariates evaluated at time $t_{ij}$ with corresponding regression coefficient vector $\beta$, which defines the population-average trajectory; $z_i(t_{ij})$ represents a vector of random-effect covariates and $b_i$ denotes the random effects, which capture subject-specific deviations from the population trajectory. $\varepsilon_{ij}$ denotes the measurement error following a normal distribution $\mathcal{N}(0, \sigma_\varepsilon^2)$. The measurement errors $\varepsilon_{ij}$ are assumed independent of $b_i$, and the longitudinal measurements $y_{ij}$ for subject $i$ conditional on $b_i$ are assumed independent across observation times.

To accommodate the sequential treatment assignment within the SMART design, we specify the true latent trajectory as

$$m_i(t_{ij}) = \beta_0 + {\beta_{X_0}}^\top X_{0i} + \beta_t t_{ij} + \beta_{V_1} min(t_{ij}, \tau) + I(t_{ij} > \tau)\beta_{V_2}(t_{ij} - \tau) + b_{0i} + b_{1i}t_{ij}.$$

Here, $X_{0i}$ represents the baseline covariates with coefficient vector $\beta_{X_0}$. Treatment effects are incorporated through piecewise cumulative exposure terms to ensure a continuous trajectory with a structural change at the decision point $\tau$: before $\tau$, only the first-stage treatment $V_1$ influences the longitudinal trajectory through a treatment-specific slope $\beta_{V_1}$ applied to cumulative exposure time $\min(t_{ij}, \tau)$; after $\tau$, the first-stage treatment effect holds constant at $\beta_{V_1}\tau$ and the second-stage treatment $V_2$ contributes through an independent slope $\beta_{V_2}$ applied to second-stage treatment exposure time $I(t_{ij} > \tau)(t_{ij} - \tau)$. The random effects $b_i$ follow a bivariate normal distribution with a covariance matrix $G$,

$$b_i = \begin{pmatrix} b_{0i} \\ b_{1i} \end{pmatrix} \sim \mathcal{N}(0, G), G = \begin{pmatrix} \sigma_{b0}^2 & \rho\, \sigma_{b0}\, \sigma_{b1} \\ \rho\, \sigma_{b0}\, \sigma_{b1} & \sigma_{b1}^2 \end{pmatrix},$$

where $b_{0i}$ represents the random intercept with standard deviation $\sigma_{b0}$ capturing between-subject variability in baseline biomarker levels, $b_{1i}$ denotes the random slope with standard deviation $\sigma_{b1}$ capturing heterogeneity in the rate of change over time, and $\rho$ refers to the correlation between the two random effects.

Under the LMM, the conditional distribution of the response vector $y_i = (y_{i0}, y_{i1}, \dots, y_{in_i})^\top$ given $b_i$ follows a multivariate normal distribution expressed as

$$y_i \mid b_i \sim \mathcal{N}\left(X_i\beta + Z_i b_i,\, \sigma_\varepsilon^2 \mathbb{I}_{n_i+1}\right),$$

where $X_i$ and $Z_i$ represent the subject-specific design matrices containing $x_i(t_{ij})$ and $z_i(t_{ij})$ across $n_i + 1$ repeated measurements, and $\mathbb{I}_{n_i+1}$ denotes the $(n_i + 1) \times (n_i + 1)$ identity matrix. The corresponding conditional density function is

$$f(y_i \mid b_i;\, \beta, \sigma_\varepsilon^2) = (2\pi\sigma_\varepsilon^2)^{-\frac{(n_i+1)}{2}} \exp\left\{-\frac{1}{2\sigma_\varepsilon^2}(y_i - X_i\beta - Z_i b_i)^\top (y_i - X_i\beta - Z_i b_i)\right\}.$$

**3.2 Survival submodel**

Let $T_i^*$ denote the true event time for patient $i$ and $C_i$ the censoring time. The observed follow-up time and event indicator are

$$T_i = \min(T_i^*, C_i),\, \delta_i = I(T_i^* \le C_i),$$

where $\delta_i = 1$ indicates an observed event and $\delta_i = 0$ indicates right-censoring.

In the SMART setting, treatment may change at the decision point $\tau$. Accordingly, the time-to-event process is specified through a relative risk model

$$h_i(t|X_{0i}, b_i;\ V_1, V_2) = h_0(t)\exp\{\eta_i(t|X_{0i}, b_i;\ V_1, V_2)\}.$$

We specify the baseline hazard following a Weibull distribution $h_0(t; \lambda_0, \kappa) = \lambda_0\, \kappa\, t^{\kappa-1}$ with scale parameter $\lambda_0$ and shape parameter $\kappa$, however, the methods presented here can be used with other baseline hazard distributions. A parametric baseline hazard is adopted to reduce the dimensionality of the baseline component, which lowers computational burden in the presence of random-effect integration and improves estimation stability in finite samples. The piecewise time-dependent linear predictor is defined as

$$\eta_i(t|X_{0i}, b_i;\ V_1, V_2) = \gamma_{X_0}{}^{\top} X_{0i} + \gamma_1{}^{\top} w_{1i} min(t, \tau) + I(t > \tau)\gamma_2{}^{\top} w_{2i}(t - \tau) + \alpha m_i(t),$$

where $X_{0i}$ represents the same baseline covariates used in the longitudinal submodel with survival-specific coefficient vector $\gamma_{X_0}$, $w_{1i}$ collects the first-stage treatment indicators that are active from baseline through $\tau$ with coefficient vector $\gamma_1$ (e.g. $V_A$), $w_{2i}$ collects the second-stage treatment indicators nested within the first-stage treatment that become active after $\tau$ with coefficient vector $\gamma_2$ (e.g. $V_{AA}$), and $m_i(t)$ refers to the current value of the latent longitudinal outcome defined in Section 3.1 with the current-value association parameter $\alpha$. As in the longitudinal submodel, treatment effects enter the hazard through cumulative exposure time rather than as instantaneous shifts at the decision point, reflecting the assumption that the impact of treatment on survival accumulates gradually with duration of exposure. However, this framework also allows instantaneous treatment effects at the decision point, and the implementation depends on the clinical context and assumptions. The current-value parameterization is adopted because it provides a direct clinical interpretation of the association between the contemporaneous latent biomarker level and the instantaneous hazard of the event.

Then, the cumulative hazard for patient $i$ under a treatment sequence $d = (V_1, V_2)$ is

$$H_{V_1,V_2}(t \mid X_{0i}, b_i) = \int_0^{\min(t,\tau)} h(u \mid X_{0i}, b_i;\ V_1)\, du + I(t > \tau) \int_\tau^t h(u \mid X_{0i}, b_i;\ V_1, V_2)\, du.$$

The corresponding survival function is

$$S_{V_1,V_2}(t \mid X_{0i}, b_i) = \exp\{-H_{V_1,V_2}(t \mid X_{0i}, b_i)\},$$

and the conditional density of the observed survival data $(T_i, \delta_i)$ is

$$f(T_i, \delta_i \mid X_{0i}, b_i; V_1, V_2) = [h_i(T_i \mid X_{0i}, b_i; V_1, V_2)]^{\delta_i} S_{V_1,V_2}(T_i \mid X_{0i}, b_i).$$

## 4. Maximum likelihood estimation

### 4.1 Joint likelihood construction

For subject $i$, let $y_i = (y_{i0}, y_{i1}, \dots, y_{in_i})^\top$ denote the longitudinal measurements observed at times $t_i = (t_{i0}, t_{i1}, \dots, t_{in_i})^\top$, let $(T_i,\ \delta_i)$ denote the observed survival data, and let $\theta = (\beta, \sigma_\varepsilon^2, \lambda_0, \kappa, \gamma, \alpha, G)$ collect all unknown parameters in the joint model: the longitudinal fixed-effects coefficients $\beta$, residual variance $\sigma_\varepsilon^2$, baseline hazard parameters $\lambda_0$ and $\kappa$, survival regression coefficients $\gamma$, association parameter $\alpha$, and random-effects covariance matrix $G$. Conditional on the subject-specific random effects $b_i$, the complete-data likelihood for $(y_i, T_i, \delta_i, b_i)$ is expressed as

$$L_i^c(\theta; b_i) = f(y_i \mid b_i; \beta, \sigma_\varepsilon^2) f(T_i, \delta_i \mid X_{0i}, b_i, V_1, V_2; \lambda_0, \kappa, \beta, \gamma, \alpha) f(b_i; G).$$

Note that $\beta$ appears in the survival density because the latent trajectory $m_i(t)$, which enters the hazard through the association term $\alpha m_i(t)$, depends on the longitudinal fixed-effects coefficients.

Since the random effects $b_i$ are not directly observed, likelihood-based inference is conducted using the observed-data likelihood by integrating out the random effects:

$$L_i(\theta) = \int L_i^c(\theta; b_i) db_i = \int f(y_i \mid b_i;\ \beta, \sigma_\varepsilon^2) f(T_i, \delta_i \mid X_{0i}, b_i, V_1, V_2; \lambda_0, \kappa, \beta, \gamma, \alpha) f(b_i; G) db_i.$$

As defined in Section 3.1, the random-effect density is

$$f(b_i; G) = (2\pi)^{-q/2} \mid G \mid^{-1/2} \exp\{-\frac{1}{2} b_i^\top G^{-1} b_i\},$$

where $q = 2$ denotes the dimension of $b_i$. The longitudinal and survival processes are linked through the shared random effects $b_i$, and $\alpha$ quantifies the strength of association between the latent longitudinal trajectory and the event hazard. When $\alpha = 0$, the joint model reduces to separate longitudinal and survival submodels. Assuming independence across subjects, the full observed-data likelihood is $L(\theta) = \prod_{i=1}^{N} L_i(\theta)$, and the corresponding log-likelihood is $\ell(\theta) = \sum_{i=1}^{N} \log L_i(\theta)$.

**4.2 Score representation and posterior form**

The objective of maximum likelihood estimation is to maximize the observed-data log-likelihood $\ell(\theta)$introduced in Section 4.1. The corresponding observed-data score function is given by

$$U(\theta) = \frac{\partial \ell(\theta)}{\partial \theta} = \sum_{i=1}^{N} U_i(\theta), \text{where } U_i(\theta) = \frac{\partial}{\partial \theta} \log L_i(\theta).$$

A direct evaluation of $U_i(\theta)$ is challenging since $L_i(\theta) = \int L_i^c(\theta; b_i)\, db_i$ involves the integral over the random effects without a closed-form expression. However, the observed-data score function can be rewritten as

$$U_i(\theta) = \frac{1}{L_i(\theta)} \frac{\partial}{\partial \theta} \int L_i^c(\theta; b_i)\, db_i = \frac{1}{L_i(\theta)} \int \frac{\partial}{\partial \theta} L_i^c(\theta; b_i)\, db_i.$$

Let

$$\mathcal{J}(\theta, b_i) = \frac{\partial}{\partial \theta} \log L_i^c(\theta; b_i)$$

denote the complete-data score function. Noting that

$$\frac{\partial}{\partial \theta} L_i^c(\theta; b_i) = L_i^c(\theta; b_i)\, \mathcal{J}(\theta, b_i),$$

we obtain

$$U_i(\theta) = \int \mathcal{J}(\theta, b_i) \frac{L_i^c(\theta; b_i)}{L_i(\theta)} \, db_i.$$

The ratio of the complete-data likelihood to the observed-data likelihood

$$\frac{L_i^c(\theta; b_i)}{L_i(\theta)} = f(b_i \mid y_i, T_i, \delta_i; \theta) = \frac{f(y_i \mid b_i; \beta, \sigma_\varepsilon^2) \, f(T_i, \delta_i \mid X_{0i}, b_i; \lambda_0, \kappa, \beta, \gamma, \alpha) \, f(b_i; G)}{L_i(\theta)}$$

is the posterior density of the random effects given the observed data. Therefore, the observed-data score can be expressed as the posterior expectation of the complete-data score,

$$U_i(\theta) = \int \mathcal{J}(\theta, b_i) \, f(b_i \mid y_i, T_i, \delta_i; \theta) \, db_i = E[\mathcal{J}(\theta, b_i) \mid y_i, T_i, \delta_i; \theta].$$

This representation shows that the complete-data score $\mathcal{J}(\theta, b_i)$ measures how the log-likelihood changes with $\theta$ when the random effects are known, while the observed-data score can be treated as an average of the complete-data score over the uncertainty in $b_i$. As a result, evaluation of the observed-data score reduces to numerical integration of $\mathcal{J}(\theta, b_i)$ over the posterior distribution $f(\, b_i \mid y_i, T_i, \delta_i; \theta \,)$.

**4.3 Numerical evaluation of the survival integral**

Evaluation of the survival component requires computation of the cumulative hazard introduced in Section 3.2. Under the joint modeling framework, this integral does not have a closed-form expression because the hazard depends on the time-varying longitudinal trajectory through the shared random effects $b_i$. We approximate it using a 15-point Gauss–Kronrod quadrature rule, which evaluates the cumulative hazard integrand at 15 predetermined nodes and combines the results as a weighted sum.[42] Let $\{(\chi_{GK}, \varpi_{GK})\}_{GK=1}^{15}$ denote the nodes and weights of the standard 15-point Gauss–Kronrod rule on the reference interval $[-1,1]$. For an arbitrary interval $[a, b]$, the integral of a function $f(t)$ is approximated by

$$\int_a^b f(t)dt \approx \sum_{GK=1}^{15} \tilde{\varpi}_{GK} f(t_{GK})$$

with an affine transformation $t_{GK} = \frac{b-a}{2}(\chi_{GK} + 1) + a$ and corresponding rescaled weights $\tilde{\varpi}_{GK} = \frac{b-a}{2}\varpi_{GK}$ mapping $[a, b]$ to $[-1,1]$. To accommodate the piecewise structure at $\tau$, the cumulative hazard is evaluated as the sum of two components: the first integrates over $[0, \min(T_i, \tau)]$ using the first-stage linear predictor, and the second integrates over $(\tau, T_i]$ using the full treatment-sequence linear predictor.

**4.4 Approximation of the random-effect integral**

The observed-data likelihood also requires integration over the random effects. Unlike the survival integral in Section 4.3, this is a $q$-dimensional integral over the random-effect vector $b_i$, and we use pseudo-adaptive Gauss–Hermite quadrature to approximate it.[43]

Standard Gauss–Hermite quadrature approximates integrals over $(-\infty, +\infty)$ by a weighted sum of the integrand evaluated at fixed nodes. Let $\{(\varphi_k, \varpi_k)\}_{k=1}^{K}$ denote the nodes and weights of the univariate $K$-point Gauss–Hermite rule. For our random effects with dimension $q = 2$, the quadrature consists of $K^2$ nodes indexed by $r = (r_1, r_2)$ with each $r_\iota \in \{1, \ldots, K\}$. The corresponding nodes and weights are given by $\varphi_r = (\varphi_{r_1}, \varphi_{r_2})^\top$ and $\varpi_r = \prod_{\iota=1}^{2} \varpi_{r_\iota}$. Standard Gauss–Hermite quadrature places nodes symmetrically around zero, which works well when the integrand is concentrated near zero. However, the posterior distribution of $b_i$ is typically shifted away from zero toward subject-specific values and concentrated in a region that standard nodes may not adequately cover.

The fully adaptive Gauss–Hermite rule addresses this problem by recentering and rescaling the quadrature nodes at every iteration of the optimization algorithm using the posterior mode $\hat{b}_i(\theta)$ and the corresponding curvature matrix. Although this improves numerical efficiency, it requires repeated calculation of subject-specific posterior modes and Hessians throughout optimization, which is computationally expensive in joint models. To reduce this burden, we adopt a pseudo-adaptive strategy that constructs the subject-specific transformation once using information from an initial LMM fit at the start of optimization. Let

$$\tilde{b}_i = \arg\max_b \; log\{f(\, y_i \mid b; \tilde{\theta}_y \,)f(b;\; \tilde{G})\}$$

denote the empirical Bayes estimate of the random effects under the longitudinal model, where $\tilde{\theta}_y$ and $\tilde{G}$ represent the corresponding maximum likelihood estimates. Let

$$\widetilde{\mathcal{H}}_i = -\frac{\partial^2}{\partial b\; \partial b^\top}\{log\, f(\, y_i \mid b; \tilde{\theta}_y \,) + log\, f(b; \tilde{G})\}\,|_{b=\tilde{b}_i}$$

denote the negative Hessian matrix of the log-posterior random effects density evaluated at $\tilde{b}_i$. The pseudo-adaptive quadrature nodes are then defined by

$$b_{i,r} = \tilde{b}_i + \sqrt{2}\, \varrho_i^{-1}\varphi_r,$$

where $\varrho_i$ denotes the Cholesky factor of $\widetilde{\mathcal{H}}_i$, so that $\varrho_i^{-1}$ rescales the standard nodes to match the local curvature of the posterior. This transformation centers the quadrature nodes at the empirical Bayes mode and scales them by the inverse Cholesky factor of the curvature matrix. As a result, nodes are placed in the region with highest expected posterior mass. The subject-specific observed-data likelihood is then approximated by

$$L_i(\theta) \approx \sum_r \widetilde{\varpi}_{i,r}\, L_i^c(\theta; b_{i,r}),$$

where $\widetilde{\varpi}_{i,r} = \pi^{-q/2}|\varrho_i^{-1}|\varpi_r$ denotes the transformed Gauss–Hermite weights associated with node $b_{i,r}$. That is, the multidimensional integral over the random effects is replaced by a finite weighted sum of complete-data likelihood contributions evaluated at subject-specific quadrature nodes.

### 4.5 Gradient evaluation

To facilitate gradient-based optimization of the likelihood and variance inference, the subject-specific observed-data score function $U_i(\theta)$ in Section 4.2 is approximated using the same pseudo-adaptive Gauss–Hermite quadrature rule. Using the transformed quadrature nodes $b_{i,r}$ and corresponding quadrature weights $\tilde{\varpi}_{i,r}$ defined in Section 4.4, the subject-specific score takes the form

$$U_i(\theta) \approx \sum\nolimits_r \Omega_{i,r}(\theta)\, \mathcal{J}(\theta, b_{i,r}),$$

where

$$\Omega_{i,r}(\theta) = \frac{\tilde{\varpi}_{i,r}\, L_i^c(\theta; b_{i,r})}{\sum_r \tilde{\varpi}_{i,r}\; L_i^c(\theta; b_{i,r})}$$

denotes the normalized posterior weight associated with the $r$th transformed quadrature node, and the denominator is the pseudo-adaptive Gauss–Hermite approximation to the observed-data likelihood $L_i(\theta)$ introduced in Section 4.4. The approximated score is thus a weighted average of complete-data score contributions evaluated at the transformed quadrature nodes.

### 4.6 Likelihood maximization

The quadrature-approximated log-likelihood is maximized using the limited-memory Broyden–Fletcher–Goldfarb–Shanno algorithm with bound constraints (L-BFGS-B), a quasi-Newton method that accumulates gradient information from recent iterations to construct a low-memory approximation of the inverse Hessian.[44] This approach is well suited to the present setting, where the likelihood and score are available through quadrature approximation but second derivatives are not available in closed form. Finite bounds are imposed on selected parameters to avoid numerically degenerate configurations, such as near-singular covariance matrices and extreme survival parameters that may cause overflow in the survival likelihood. In addition, variance parameters ($\sigma_\varepsilon^2$, $\sigma_{b0}^2$, $\sigma_{b1}^2$) and baseline hazard parameters ($\lambda_0, \kappa$) are optimized on the logarithmic scale to ensure positivity, and the random-effect correlation, $\rho$, is optimized after Fisher $z$-transformation to enforce $|\rho| < 1$. Within L-BFGS-B, convergence is monitored using the projected gradient, which sets any component that would move the corresponding parameter outside the allowable range to zero. In this study, the optimizer is implemented with a projected-gradient tolerance of $10^{-3}$, and convergence is further verified post-optimization using the relative projected gradient with a threshold of $10^{-3}$.

To improve numerical stability and reduce computational burden, initial values, $\theta^{(0)}$, are obtained from separate submodel fits. The longitudinal parameters $\left(\beta^{(0)}, \sigma_\varepsilon{}^{(0)}, G^{(0)}\right)$ are initialized from the LMM fitted to the longitudinal data. This LMM also provides the empirical Bayes estimates and curvature matrices used to construct the pseudo-adaptive Gauss–Hermite nodes in Section 4.4. The survival regression coefficients $\gamma^{(0)}$ are initialized from the relative risk model without the survival-longitudinal association. Initial values for the baseline hazard parameters $\left(\lambda_0{}^{(0)}, \kappa^{(0)}\right)$ are obtained using the Breslow estimator of the baseline cumulative hazard from the relative risk model fit. The survival-longitudinal association parameter is initialized at $\alpha^{(0)} = 0$.

**4.7 Uncertainty estimation**

The variance-covariance matrix of the parameter estimates is obtained using the observed Fisher information matrix based on the quadrature-approximated log-likelihood $\tilde{\ell}(\theta)$.[38] The observed Fisher information matrix is defined as the negative Hessian of the log-likelihood

$$\mathcal{I}(\hat{\theta}) = -\left.\frac{\partial^2 \tilde{\ell}(\theta)}{\partial\theta\ \partial\theta^\top}\right|_{\theta=\hat{\theta}},$$

where $\hat{\theta}$ denotes the converged maximum likelihood parameter estimates. Since the Hessian of $\tilde{\ell}(\theta)$ does not have a closed-form expression, $\mathcal{J}(\hat{\theta})$ is computed numerically at $\hat{\theta}$ using finite-difference approximations. The estimated variance-covariance matrix is defined as $\widehat{\text{Var}}(\hat{\theta}) = \mathcal{J}(\hat{\theta})^{-1}$, and standard errors are obtained as the square roots of its diagonal elements. For parameters optimized on transformed scales, standard errors on the natural scale are obtained using the delta method.

**5. Regimen-specific causal inference**

**5.1 Causal identification**

In the DTR framework, causal effects are defined as contrasts of the mean counterfactual outcomes, where a counterfactual outcome represents the outcome that would have been observed if treatment had been hypothetically assigned so that all patients followed a specified DTR $g$.[45,46] In the presence of censoring, causal contrasts should reflect the counterfactual outcome that would have been observed without censoring to avoid selection bias induced by restricting to uncensored subjects.

For a DTR $g = (V_1^g, V_1^g, V_2^g)$, where $V_1^g$ denotes the first-stage treatment that responders continue and $V_2^g$ denotes the alternative second-stage treatment for non-responders, let $T_i^g$ denote the counterfactual event time that would be observed if subject $i$ follows regimen $g$. In this study, the primary causal estimands are the regimen-specific survival function, $S_g(t) = \Pr(T_i^g > t)$, and the restricted mean survival time (RMST), $\Psi_g(t^*) = E[\min(T_i^g, t^*)]$, where $t^*$ denotes a pre-specified time horizon. Thus, a causal effect between two regimens $g$ and $g'$ is defined as $S_g(t) - S_{g'}(t)$ or $\Psi_g(t^*) - \Psi_{g'}(t^*)$. Let $R_i^g = I\{T_i^g \geq \tau, y_i^g(0) - y_i^g(\tau) \geq c\}$ denote the counterfactual response indicator at the decision time $\tau$ following regimen $g$. Identification of $S_g(t)$ and $\Psi_g(t^*)$, as well as their contrasts across DTRs, from the observed two-stage SMART data requires the following assumptions: [47]

(1) Consistency: if subject $i$ follows the treatment sequence specified by $g$, then the observed outcomes equal the corresponding counterfactual outcomes. That is, if $V_{1i} = V_1^g$, then $T_i = T_i^g$ for $T_i < \tau$; if $V_{1i} = V_1^g$, $T_i \geq \tau$, $C_i \geq \tau$, and $V_{2i} = V_2^g(R_i)$, then $(T_i, R_i) = (T_i^g, R_i^g)$.

(2) Positivity: each treatment decision specified by regimen $g$ occurs with positive probability among subjects eligible for that decision under the trial design. That is,

$$\mathrm{P}(V_{1i} = V_1^g \mid X_{0i}) > 0;\ \mathrm{P}(V_{2i} = V_2^g(R_i) \mid X_{0i}, V_{1i} = V_1^g, R_i = 0, T_i \geq \tau, C_i \geq \tau) > 0.$$

(3) Sequential conditional exchangeability: conditional on the observed history, treatment assignment at each stage is independent of the relevant future counterfactual outcomes. That is,

$$V_{1i} \perp (T_i^g, R_i^g) \mid X_{0i}; V_{2i} \perp T_i^g \mid X_{0i}, V_{1i}, R_i, T_i \geq \tau, C_i \geq \tau.$$

(4) Conditional independent censoring: at each time $t$ during follow-up, among subjects who remain event-free and uncensored, the censoring process is independent of the counterfactual event time under regimen $g$ conditional on the observed treatment and covariate history. Let $y_i(t^-)$ denote the longitudinal measurements up to time $t$, then

$$C_i \perp T_i^g \mid X_{0i}, V_{1i}, R_i, V_{2i}, y_i(t^-), T_i \geq t, C_i \geq t.$$

By conditioning on the longitudinal history $y_i(t^-)$, this assumption accommodates censoring that depends on the biomarker trajectory, which is more realistic in oncology settings where dropout may be associated with disease progression.

In a SMART, sequential exchangeability holds by design because treatment is randomized at each decision stage, and positivity holds because the randomization probabilities for the embedded DTRs are known and strictly positive. Consistency holds when the treatment options and decision rule are well defined. The conditional independent censoring assumption is not guaranteed by randomization and remains an untestable assumption. However, the joint model mitigates this concern by conditioning on the longitudinal biomarker history, which accounts for a key source of informative dropout in oncology settings.

### 5.2 Parametric g-formula

The g-formula is a causal identification and estimation method for regimen-specific outcomes under DTRs, which can adjust for time-varying confounders affected by prior treatment in observational studies.[46,47] In low-dimensional settings, the g-formula can be evaluated nonparametrically by summing over all possible post-baseline histories. However, this becomes infeasible in the presence of continuous covariates or high-dimensional histories. Young et al. described a general parametric g-formula implementation in an observational discrete-time longitudinal setting, where the unknown conditional distributions were replaced by fitted parametric models and the resulting integral over all possible post-baseline histories was approximated by Monte Carlo simulation.[48]

We apply the parametric g-formula within the joint longitudinal-survival model introduced in Sections 3-4. In the present two-stage SMART, this general history integral simplifies because the post-baseline treatment adaptation depends only on the binary response indicator at one fixed decision point τ. As a result, the regimen-specific survival function can be evaluated as a response-weighted mixture of treatment-sequence-specific survival functions.

Conditional on baseline covariates $X_{0i}$, random effects $b_i$, and the regimen-specific first-stage treatment $V_1^g$, the LMM determines the distribution of the biomarker reduction from baseline to the decision time as

$$y_{i0}^g - y_{i\tau}^g \mid (X_{0i}, b_i, V_1^g) \sim N\left(m_i^g(0) - m_i^g(\tau), 2\sigma_\varepsilon^2\right),$$

where $m_i^g(t) = m_i(t;\ X_{0i}, b_i, V_1^g)$ represents the latent mean trajectory evaluated at time $t \le \tau$ under the first-stage treatment specified by $g$. Therefore, the regimen-specific response probability is defined as

$$p_g(X_{0i}, b_i) = \mathrm{P}(R_i^g = 1 \mid X_{0i}, V_1^g, b_i, T_i^g \ge \tau) = \Phi\left(\frac{m_i^g(0) - m_i^g(\tau) - c}{\sqrt{2}\sigma_\varepsilon}\right),$$

where $\Phi(\cdot)$ denotes the standard normal cumulative distribution function. We use the normal cumulative distribution function here because the response is defined solely through the longitudinal biomarker, which follows a normal distribution under the LMM. When response is defined by a compound indicator incorporating multiple clinical criteria, a logistic regression model for the response probability can be employed without altering the remainder of the g-formula framework. Subjects who are censored or experience the event before $\tau$ do not receive a second-stage treatment and therefore do not contribute to the response probability calculation. However, similar to the responders, they are included in the regimen-specific survival estimation for every DTR sharing their observed first-stage treatment, as their pre-$\tau$ outcomes are common across those regimens. Then, the regimen-specific survival function is expressed as

$$S_g^{JM}(t \mid X_{0i}, b_i; \theta) = p_g(X_{0i}, b_i)\, S_{V_1^g, V_2^g(R=1)}(t \mid X_{0i}, b_i; \theta) + \{1 - p_g(X_{0i}, b_i)\}\, S_{V_1^g, V_2^g(R=0)}(t \mid X_{0i}, b_i; \theta),$$

where $S_{(V_1,V_2)}(t \mid X_{0i}, b_i) = \exp\{-H_{(V_1,V_2)}(t \mid X_{0i}, b_i)\}$ denotes the survival function for treatment sequence $d = (V_1, V_2)$ with the piecewise cumulative hazard defined in Section 3.2. The corresponding conditional RMST is

$$\Psi_g^{JM}(t^* \mid X_{0i}, b_i; \theta) = \int_0^{t^*} S_g^{JM}(u \mid X_{0i}, b_i; \theta)\, du,$$

and the integral over time is evaluated numerically by the trapezoidal rule on a prespecified grid over $[0, t^*]$.[49] At each grid point, the survival function is obtained from the cumulative hazard, which is evaluated using the 15-point Gauss–Kronrod quadrature described in Section 4.3.

To obtain marginal regimen-specific values, we integrate the conditional functions over the random-effect distribution and average over the empirical baseline covariate distribution of the $N$ study subjects. Unlike observational settings where the g-formula requires Monte Carlo simulation to integrate over the full post-baseline covariate history, the joint model reduces the integration to the random-effect space, which can be evaluated numerically using standard Gauss–Hermite quadrature. Therefore, the marginal regimen-specific survival probability and RMST are expressed as

$$S_g^{JM}(t;\theta) = \frac{1}{N}\sum_{i=1}^{N}\int S_g^{JM}(\, t \mid X_{0i}, b; \theta\,)\, f(b; G)\, db$$

and

$$\Psi_g^{JM}(t^*;\theta) = \frac{1}{N}\sum_{i=1}^{N}\int \Psi_g^{JM}(\, t^* \mid X_{0i}, b; \theta\,)\, f(b; G)\, db,$$

where $f(b; G)$ represents the $q$-dimensional random-effect density defined in Section 4.1. Accordingly, all subjects contribute to the estimation of $S_g^{JM}(t)$ or $\Psi_g^{JM}(t^*)$ through the common joint-model parameters and standardization over the observed baseline covariate distribution, even though some of them do not actually follow DTR $g$. After obtaining the maximum likelihood estimate $\hat{\theta}$, the plug-in estimators $\hat{S}_g^{\mathrm{JM}}(t;\hat{\theta})$ and $\hat{\Psi}_g^{\mathrm{JM}}(t^*;\hat{\theta})$ are obtained by evaluating these expressions at $\hat{\theta}$.

### 5.3 IPTW Kaplan–Meier estimator

As a comparator to the proposed joint-model framework, we consider an inverse probability of treatment weighted (IPTW) Kaplan–Meier estimator based on the known SMART randomization probabilities.[22,30] This nonparametric estimator is robust to outcome model misspecification. For a DTR $g$, the IPTW estimator of the regimen-specific survival function is defined as

$$\hat{S}_g^{\text{IPTW}}(t) = \prod_{t_{(s)} \leq t} \left[ 1 - \frac{\sum_{i=1}^{N} W_i\,(g)\, I(T_i = t_{(s)}, \delta_i = 1)}{\sum_{i=1}^{N} W_i\,(g)\, I\left(T_i \geq t_{(s)}\right)} \right].$$

where $t_{(1)} < \cdots < t_{(s)}$ denote the distinct observed event times and $W_i(g)$ represents the subject-specific IPTW weight. In our two-stage SMART introduced in Section 2 with $P_1 = P_2 = 0.5$, $W_i(g) = 0$ for subjects whose observed treatment sequence is incompatible with regimen $g$; $W_i(g) = 1/P_1 = 2$ for subjects who respond to their first-stage treatment, experience the event before $\tau$, or are censored before $\tau$; and $W_i(g) = 1/(P_1 P_2) = 4$ for nonresponders whose observed second-stage treatment matches the nonresponder arm of regimen $g$. Responders and subjects who experience the event or are censored before $\tau$ contribute to both DTRs sharing their observed first-stage treatment. Since $\hat{S}_g^{\text{IPTW}}(t)$ is a step function with jumps at the observed event times $t_1 < \cdots < t_s$, the regimen-specific RMST up to horizon $t^*$ is obtained as a sum of rectangular contributions over the observed event times:

$$\hat{\Psi}_g^{\text{IPTW}}(t^*) = \sum\nolimits_{s:\, t_s \leq t^*} \hat{S}_g^{\text{IPTW}}\,(t_s)(\min\,(t_{s+1}, t^*) - t_s),\, t_0 = 0.$$

**5.4 Optimal DTR identification**

A central goal of SMARTs is to identify the optimal DTR among all embedded regimens. Let $\mathcal{G} = \{g_1, \dots, g_M\}$ denote the set of embedded DTRs, and let $Q_g$ denote the regimen-specific estimand used to rank regimens (such as $S_g(t)$ and $\Psi_g(t^*)$ in this study). Assuming higher $Q_g$ indicates more favorable outcomes, as is common in survival studies where the event is often a poor outcome, the optimal DTR refers to the regimen with the largest $Q_g$ denoted as $g^* = \arg\max_{g \in \mathcal{G}} Q_g$. However, the DTR with the largest point estimate may not be truly optimal due to sampling variability. In a SMART with $M$ embedded DTRs, exhaustive pairwise comparisons require $M(M-1)/2$ tests, and standard multiplicity corrections such as Bonferroni can become overly conservative as $M$ increases. Multiple comparisons with the best (MCB) addresses this problem by constructing a confidence set of regimens that are statistically indistinguishable from the optimal after controlling for multiplicity.[50] Chao et al. extended the original MCB algorithm to survival outcomes in the SMART setting.[37]

For each regimen $g \in \mathcal{G}$, MCB is formulated through the contrasts $\Delta_g = Q_g - \max_{g' \neq g} Q_{g'}$, which compares $g$ to its strongest competitor $g' \in \mathcal{G}$. $\Delta_g > 0$ implies $g$ is uniquely optimal, $\Delta_g = 0$ corresponds to a tie with the best, and $\Delta_g < 0$ implies $g$ is inferior to at least one competing regimen. Let $\hat{Q}_g$ denote the estimated value of $Q_g$, and let $\hat{Q}_{\mathcal{G}} = \left(\hat{Q}_{g_1}, \dots, \hat{Q}_{g_M}\right)^{\top}$ denote the vector of estimated regimen values with covariance matrix $\hat{\Sigma}_{\mathcal{G}}$. For each pair $(g, g') \in \mathcal{G}$, the standard error of the pairwise contrast $\hat{Q}_g - \hat{Q}_{g'}$ is defined as

$$\widehat{\text{se}}_{gg'} = \sqrt{\text{Var}\left(\hat{Q}_{\mathcal{G},g} - \hat{Q}_{\mathcal{G},g'}\right)} = \sqrt{\hat{\Sigma}_{\mathcal{G},gg} + \hat{\Sigma}_{\mathcal{G},g'g'} - 2\hat{\Sigma}_{\mathcal{G},gg'}},$$

where $\hat{\Sigma}_{\mathcal{G},gg}$ denotes the estimated variance of $\hat{Q}_g$, $\hat{\Sigma}_{\mathcal{G},g'g'}$ denotes the estimated variance of $\hat{Q}_{g'}$, and $\hat{\Sigma}_{\mathcal{G},gg'}$ denotes the estimated covariance between $\hat{Q}_g$ and $\hat{Q}_{g'}$. Then, the MCB confidence set at level $1 - \varsigma$ is given by

$$\hat{\mathcal{G}}^*_{\varsigma} = \left\{ g \in \mathcal{G} : \hat{Q}_g \geq \max_{g' \neq g} \left(\hat{Q}_{g'} - \mathcal{D}_{g,\varsigma} \widehat{\text{se}}_{gg'}\right) \right\},$$

where $\mathcal{D}_{g,\varsigma}$ represents the multiplicity-adjusted threshold that controls the family-wise error rate across all $M$ regimens at level $1-\varsigma$. Equivalently, the MCB margin for regimen $g$, $\mathbb{M}_g = \mathcal{D}_{g,\varsigma} - \max_{g' \neq g} \frac{\hat{Q}_{g'} - \hat{Q}_g}{\widehat{se}_{gg'}}$, quantifies how far it sits from the boundary of $\hat{\mathcal{G}}_\varsigma^*$, with $g \in \hat{\mathcal{G}}_\varsigma^*$ if $\mathbb{M}_g \geq 0$. If $\hat{\mathcal{G}}_\varsigma^*$ contains a single regimen, it is reported as the estimated optimal DTR; otherwise, it represents the set of DTRs statistically indistinguishable from the optimal at level $\varsigma$.

Implementation of MCB requires estimates of $\hat{Q}_g$, $\hat{\Sigma}_{\mathcal{G}}$ and $\mathcal{D}_{g,\varsigma}$, which differ between the joint-model framework and the IPTW estimator. For the joint-model framework, as defined in Section 5.2, $\hat{Q}_g^{\mathrm{JM}}$ is the plug-in parametric g-formula estimator evaluated at the maximum likelihood estimate $\hat{\theta}$, such as $\hat{S}_g^{\mathrm{JM}}(t)$ and $\hat{\Psi}_g^{\mathrm{JM}}(t^*)$. Then, we generate $N_{\mathrm{JM}}$ draws indexed by $n_{\mathrm{JM}} = 1, \dots, N_{\mathrm{JM}}$, sample $\tilde{\theta}_{n_{\mathrm{JM}}} \sim N(\hat{\theta}, \widehat{\mathrm{Var}}(\hat{\theta}))$ and recompute the regimen-specific summaries $\tilde{Q}_{g,n_{\mathrm{JM}}}^{\mathrm{JM}}$ under each draw and compute the sample covariance matrix of the $N_{\mathrm{JM}}$ simulated vectors to obtain $\hat{\Sigma}_{\mathcal{G}}^{JM}$. Since each draw produces estimates for all regimens from the same fitted joint model, the resulting covariance matrix retains both within- and between-first-stage dependence. In our two-stage SMART with four embedded regimens, we use the full covariance matrix

$$\hat{\Sigma}_{\mathcal{G}}^{JM} = \begin{pmatrix} \hat{\Sigma}_{AA} & \hat{\Sigma}_{AB} \\ \hat{\Sigma}_{BA} & \hat{\Sigma}_{BB} \end{pmatrix} = \begin{pmatrix} \hat{\Sigma}_{AAC,AAC} & \hat{\Sigma}_{AAC,AAD} & \hat{\Sigma}_{AAC,BBC} & \hat{\Sigma}_{AAC,BBD} \\ \hat{\Sigma}_{AAD,AAC} & \hat{\Sigma}_{AAD,AAD} & \hat{\Sigma}_{AAD,BBC} & \hat{\Sigma}_{AAD,BBD} \\ \hat{\Sigma}_{BBC,AAC} & \hat{\Sigma}_{BBC,AAD} & \hat{\Sigma}_{BBC,BBC} & \hat{\Sigma}_{BBC,BBD} \\ \hat{\Sigma}_{BBD,AAC} & \hat{\Sigma}_{BBD,AAD} & \hat{\Sigma}_{BBD,BBC} & \hat{\Sigma}_{BBD,BBD} \end{pmatrix},$$

where $\hat{\Sigma}_{AA}$ and $\hat{\Sigma}_{BB}$ denote the covariance matrix for DTRs starting with first-stage treatment A and B, respectively; $\hat{\Sigma}_{AB}$ and $\hat{\Sigma}_{BA}$ capture the covariance from DTRs with different first-stage treatments.

For the IPTW estimator, $\hat{Q}_g^{\text{IPTW}}$ is obtained based on the known SMART randomization probabilities as described in Section 5.3. $\hat{\Sigma}_{\mathcal{G}}^{\text{IPTW}}$ is estimated by subject-level nonparametric bootstrap. Let $N_{\text{IPTW}}$ denote the number of bootstrap replicates. For each replicate $n_{\text{IPTW}} = 1, \dots, N_{\text{IPTW}}$, we resample $N$ subjects with replacement from the observed dataset with their full observed treatment and outcome history. We recompute $\hat{Q}_{g,n_{\text{IPTW}}}^{\text{IPTW}}$ on each resampled dataset and use the sample covariance of the $N_{\text{IPTW}}$ bootstrap replicates as $\hat{\Sigma}_{\mathcal{G}}^{\text{IPTW}}$. For MCB, we impose a block-diagonal covariance structure

$$\hat{\Sigma}_{\mathcal{G}}^{\text{IPTW}} = \begin{pmatrix} \hat{\Sigma}_{AA} & 0 \\ 0 & \hat{\Sigma}_{BB} \end{pmatrix} = \begin{pmatrix} \hat{\Sigma}_{AAC,AAC} & \hat{\Sigma}_{AAC,AAD} & 0 & 0 \\ \hat{\Sigma}_{AAD,AAC} & \hat{\Sigma}_{AAD,AAD} & 0 & 0 \\ 0 & 0 & \hat{\Sigma}_{BBC,BBC} & \hat{\Sigma}_{BBC,BBD} \\ 0 & 0 & \hat{\Sigma}_{BBD,BBC} & \hat{\Sigma}_{BBD,BBD} \end{pmatrix},$$

and treat DTRs with different first-stage treatments as uncorrelated since only subjects whose observed data are compatible with regimen $g$ receive positive weight for that regimen.

The cutoff $\mathcal{D}_{g,\varsigma}$ depends on $\hat{\Sigma}_{\mathcal{G}}$. Let $\xi = (\xi_{g_1}, \dots, \xi_{g_M})^\top \sim N(0, \hat{\Sigma}_{\mathcal{G}})$ denote an auxiliary vector, then $\mathcal{D}_{g,\varsigma}$ satisfies $1 - \varsigma = P\left(\mathcal{D}_{g,\varsigma} \geq \max_{g' \neq g} \frac{\xi_{g'} - \xi_g}{\widehat{\text{se}}_{gg'}}\right)$. Since this probability has no closed form, we approximate it by Monte Carlo simulation. For each regimen $g \in \mathcal{G}$, we generate $N_{\text{MC}}$ independent Monte Carlo draws indexed by $n_{\text{MC}} = 1, \dots, N_{\text{MC}}$, $\xi_{n_{\text{MC}}} \sim N\left(0, \hat{\Sigma}_{\mathcal{G}}\right)$, compute the maximum pairwise standardized gap between any competitor and regimen $g$,

$$\mathbb{G}_{g,\text{n}_{\text{MC}}} = \max_{g' \neq g} \frac{\xi_{g',n_{\text{MC}}} - \xi_{g,n_{\text{MC}}}}{\widehat{\text{se}}_{g,g'}},$$

and take the $(1-\varsigma)$-quantile of $\left\{\mathbb{G}_{g,\text{n}_{\text{MC}}} : n_{\text{MC}} = 1, \dots, N_{\text{MC}}\right\}$ $as\ \mathcal{D}_{g,\varsigma}$.

## 6. Simulation study

### 6.1 Data-generating mechanism

We conducted a Monte Carlo simulation study under the two-stage SMART design described in Section 2 with decision point $\tau = 8$ and response threshold $c = 1.3$. Three sample sizes $N \in \{300,600,1200\}$ were evaluated with 1000 replications. In each replication, subjects were followed from baseline to an administrative end of follow-up at $t_{\max} = 24$. An independent random censoring mechanism was introduced as $C_i \sim \text{Exp}(0.15)$ to represent loss to follow-up during the study period.

For numerical stability, all time-dependent quantities were internally rescaled by a factor of 1/10 during estimation. The baseline hazard parameters ($\lambda_0$, $\kappa$) and all time-dependent coefficients were reported on this rescaled time, and survival outcomes ($S_g(t^*)$, $\Psi_g(t^*)$) were reported on the original scale.

We generated two baseline covariates. The first binary covariate was drawn from a Bernoulli distribution with probability 0.6, denoted as $X_{01i} \sim \text{Bernoulli}(0.6)$; the second continuous covariate was drawn from a standard normal distribution, denoted as $X_{02i} \sim \mathcal{N}(0,1)$. They were both incorporated in the longitudinal and survival submodels.

The longitudinal outcome was generated from the LMM introduced in Section 3.1. We considered two simulation scenarios to evaluate robustness to longitudinal measurement frequency. In the primary setting, biomarker measurements were collected at $t_{ij} \in \{0, 1, 2, \dots, 24\}$. In the secondary analysis, we reduced to four measurements at $t_{ij} \in \{0, 8, 16, 24\}$, representing sparse longitudinal data commonly encountered in real-world clinical studies. The true latent trajectory was specified as

$$m_i(t) = \beta_0 + \beta_{X_{01}} X_{01i} + \beta_{X_{02}} X_{02i} + \beta_t t + \beta_{V_{1i}} min(t, \tau) + I(t > \tau)\beta_{V_{2i}}(t - \tau) + b_{0i} + b_{1i} t,$$

where $\beta_{V_{1i}}$ and $\beta_{V_{2i}}$ denote the treatment-specific slopes corresponding to the first- and second-stage treatments received by patient $i$. We set $\beta_A = -0.8$, $\beta_B = -0.6$, $\beta_C = -0.7$, with treatment

D as the reference ($\beta_D = 0$). For the fixed population-average effects, we set $\beta_0 = 3.5$, $\beta_{X_{01}} = 0.5$, $\beta_{X_{02}} = 0.7$, and $\beta_t = -0.5$. Subject-specific random effects followed a bivariate normal distribution with mean 0; standard deviations $\sigma_{b0} = 0.5$, $\sigma_{b1} = 0.2$; and correlation $\rho = -0.3$. The standard deviation of measurement error was specified as $\sigma_\varepsilon = 0.5$.

Event times were generated from the hazard function described in Section 3.2 with a Weibull baseline hazard ($\lambda_0 = 0.15$, $\kappa = 2.6$). The time-dependent linear predictor following the piecewise structure of the survival model was specified as:

$$\eta_i(t) = \gamma_{X_{01}} X_{01i} + \gamma_{X_{02}} X_{02i} + \gamma_A I(V_{1i} = A) min(t, \tau) +$$
$$I(t > \tau)(t - \tau)[\gamma_{AA} I(V_{1i} = A, V_{2i} = A) + \gamma_{BB}\, I(V_{1i} = B, V_{2i} = B) + \gamma_{AC}\, I(V_{1i} = A, V_{2i} = C)$$
$$+ \gamma_{BC}\, I(V_{1i} = B, V_{2i} = C)] + \alpha m_i(t).$$

Baseline covariate effects were set as $\gamma_{X_{01}} = 0.4$ and $\gamma_{X_{02}} = 0.2$. We set $\gamma_A = -0.5$ with treatment B as the first-stage reference. We assumed that the effect of a second-stage treatment may depend on prior treatment history to reflect potential treatment interaction. Therefore, the second-stage treatment effects were quantified through four parameters with treatment $D$ as the reference within each first-stage arm. We set $\gamma_{AA} = -1.5$ and $\gamma_{BB} = -1.4$ as the contrasts of continuation versus switching to D; $\gamma_{AC} = -1.0$ and $\gamma_{BC} = -0.9$ as the contrasts of switching to C versus D. The association parameter linking the longitudinal process to survival was set to $\alpha = 0.2$. All true parameter values specified above are summarized in Table 1.

The observed event times were then obtained using the inverse transformation of the subject-specific cumulative hazard, $H_i(T_i^*)$. For each subject $i$, an independent exponential variable $E_i \sim \text{Exp}(1)$ was drawn and the true event time $T_i^*$ was defined as the solution to $H_i(T_i^*) = E_i$. For data generation, the cumulative hazard was approximated using composite Simpson's rule with 24 subintervals.[49] The inverse-transform procedure was applied sequentially under the SMART design. We first computed $H_i(t)$ under the first-stage treatment over $(0, \tau]$. If $H_i(\tau) \geq E_i$, the event occurred before $\tau$ and $T_i^*$ was the solution to $H_i(T_i^*) = E_i$ within $(0, \tau]$. Otherwise, $H_i(t)$ was recomputed under the complete treatment sequence, and $T_i^*$ was obtained as the solution to $H_i(T_i^*) = E_i$ over $(\tau, t_{max}]$. Administrative censoring was imposed at $t_{\max} = 24$, and the observed survival time was $T_i = \min(T_i^*, C_i, t_{\max})$ with event indicator $\delta_i = I(T_i^* \leq min(C_i, t_{max}))$. Longitudinal biomarker measurements were truncated at each patient's observed follow-up time $T_i$.

Under our data generating mechanism, the mean event and censor rates for initial treatments A and B at the decision point were 15.8% and 10.7%, respectively; the overall event and censor rates for all treatment sequences at the end of follow-up were 58.2% and 20.7%, respectively; and the response rate among subjects surviving to week 8 was 0.32.

**6.2 MLE implementation**

For each simulated dataset, joint-model parameters were estimated by maximum likelihood described in Section 4, with $K = 5$ nodes per random-effect dimension in the pseudo-adaptive Gauss–Hermite quadrature. The performance of MLE was summarized across replications. For each parameter in $\theta$, relative bias was computed as $(\bar{\hat{\theta}} - \theta^*)/\theta^* \times 100\%$, where $\theta^*$ represents the true parameter values and $\bar{\hat{\theta}}$ represents the average MLE; Monte Carlo standard error (MCSE) was defined as the empirical standard deviation of $\hat{\theta}$ across replications; average estimated standard error (AESE) referred to the average of the model-based standard errors from the inverse observed Fisher information matrix; root mean squared error (RMSE) was calculated as the square root of the mean squared deviation of $\hat{\theta}$ from $\theta^*$; and coverage probability (Cov%) was defined as the proportion of replications where the 95% Wald confidence interval based on AESE contained $\theta^*$.

**6.3 Regimen-specific value estimation**

We evaluated regimen-specific survival probability $S_g(t^*)$ and $\Psi_g(t^*)$ at horizons $t^* \in \{16,24\}$ using both the joint-model framework and the IPTW Kaplan–Meier estimator. For the marginal g-formula integration over random effects, we used Gauss–Hermite quadrature with $K = 3$ nodes per random-effect dimension as the marginal integrand over the full random-effects distribution is smoother. The time integral for $\Psi_g(t^*)$ was computed by the trapezoidal rule on a grid of 100 time points over $[0, t^*]$ for both estimators.

To evaluate estimation accuracy, true regimen-specific values $S_g^*(t^*)$ and $\Psi_g^*(t^*)$ were computed using the parametric g-formula under the data-generating parameter vector $\theta^*$. We employed a high-precision scheme with 5,000 Monte Carlo draws over the baseline covariate distribution, Gauss–Hermite quadrature with $K = 5$ nodes per dimension, and a grid of 500 time points.

For each DTR $g$, estimand, and horizon $t^*$ across replications, we reported relative bias, MCSE, RMSE, and point-estimate optimal DTR selection accuracy (Point%) defined as the proportion of replications in which the estimator correctly identifies the regimen with the largest point estimate. In addition to the regimen-specific values, we evaluated pairwise contrasts between DTRs. To compare efficiency between the two approaches, we assessed the relative efficiency (RE), defined as the ratio of variances between estimators ($\text{RE} = \text{MCSE}_{\text{JM}}^2/\text{MCSE}_{\text{IPTW}}^2$). To identify the optimal DTR after accounting for multiplicity, we applied the MCB defined in Section 5.4 with $\varsigma = 0.05$ at N = 300. The joint-model MCB procedure is computationally intensive because each multivariate normal draw requires recomputing regimen-specific values through the plug-in parametric g-formula; however, the resulting covariance estimates are relatively stable under the asymptotic distribution of $\hat{\theta}$. In contrast, bootstrap-based covariance estimation is computationally efficient but has high variability with resampling. Accordingly, we used 300 draws for the joint-model framework and 1000 bootstrap replicates for the IPTW estimator. MCB performance was summarized by the inclusion probability of each DTR in the estimated best set (MCB%), the average size of the best set, and the MCB margin ($\mathbb{M}$).

The multivariate normal draws and bootstrap replicates used for MCB also provided within-replication standard errors for regimen-specific values and pairwise contrasts, estimated by the standard deviation of recomputed values within each simulation replicate. AESE was defined as the average of these standard errors across replications, and Cov% was calculated as the proportion of replications where the 95% Wald confidence interval based on AESE contained the corresponding true DTR-specific or contrast value.

**6.4 Simulation results**

Across all simulation conditions, the L-BFGS-B optimizer achieved convergence in 100% of the 1000 replications under the convergence criterion defined in Section 4.6, and all reported summaries were based on the full set of replications.

Overall, the joint model generated accurate and precise parameter estimates across all simulation conditions. As shown in Table 1, estimation accuracy improved as sample size increased, with reductions in relative bias, standard errors, and RMSE, while coverage remained close to the nominal 95% level. The close agreement between AESE and MCSE across all sample sizes indicated that the variance estimator based on the observed information matrix was well-calibrated even at N=300. The longitudinal submodel was particularly stable, reflecting the substantial within-subject information provided by up to 25 repeated measurements per subject. In contrast, the survival submodel was more challenging, especially when N = 300. For example, the baseline hazard scale parameter $\lambda_0$ showed a relative bias of 11.10% when N=300. However, this was primarily driven by finite-sample effects, as the bias decreased to 2.20% when N=1200.

Under our parameter specification, (A, A, C) was the theoretically optimal DTR with the highest true regimen-specific values across all four estimands (Table 2). The ordering among non-optimal regimens was generally consistent, although (B, B, C) slightly exceeded (A, A, D) for $S(24)$. As shown in Table 2 for N = 300, both the joint-model framework and the IPTW Kaplan–Meier estimator provided stable and essentially unbiased estimation of DTR-specific survival outcomes at horizons $t^* \in \{16,24\}$. However, the joint-model framework consistently produced lower variability with RE ranging from 0.45 to 0.89. This efficiency advantage stemmed from the parametric structure of the joint model and the information from the longitudinal data. Meanwhile, the AESEs for the joint-model framework were obtained from propagating multivariate normal draws of $\hat{\theta}$ through the plug-in parametric g-formula, which may slightly underestimate sampling variability due to reliance on asymptotic approximations. As a result, joint-model coverage fell slightly below the nominal 95% level while bootstrap-based IPTW coverage remained closer to 95%.

Additionally, the joint-model framework identified (A, A, C) as the optimal DTR more consistently than the IPTW estimator based on point estimates (Table 2). The difference was largest for $\Psi(16)$ and $S(16)$, where the joint-model framework selected the true optimum in 99.5% and 99.4% of replications, compared with 71.0% and 85.3% for IPTW. The MCB results reinforced this distinction. The difference was most pronounced for $\Psi(16)$, where (A, A, D), (B, B, C), and (B, B, D) were included in the joint-model best set for 9.0%, 22.0%, and 2.6% of replications, compared with 94.6%, 63.5%, and 40.6% under the IPTW estimator. Since MCB depends on both point estimates and their uncertainty, the joint-model framework's smaller standard errors led to stronger separation between the best regimen and its competitors. Notably, the joint-model framework included (B, B, C) in the best set more often than (A, A, D), despite (A, A, D) having larger true values for $\Psi(16)$, $\Psi(24)$, and $S(16)$.

Table 3 provides a detailed comparison of pairwise DTR contrasts and reveals several additional patterns beyond those observed for regimen-specific estimates. First, relative bias was inflated for contrasts with small true differences. For example, the true contrast for $S(24)$ between (A, A, D) and (B, B, C) was $-0.0096$, and the relative bias over 30% corresponded to only around 0.0031 absolute bias. Second, the efficiency advantage of the joint-model framework was more pronounced for contrasts than for DTR-specific estimates, especially for contrasts between DTRs sharing the same first-stage treatment. Unlike the IPTW estimator, which relies on weighted empirical averages computed separately for each DTR, regimen-specific estimates in the joint model are computed from the same fitted parameters. For DTRs sharing the same first-stage treatment, the corresponding first-stage parameter uncertainty affects both regimens identically and cancels in the contrast. This explains why (B, B, C) was included in the joint-model best set more often than (A, A, D) in Table 2. Specifically, the largest gain was observed for $\Psi(16)$ between (A, A, C) and (A, A, D), where the joint-model variance was reduced by over 90% relative to the IPTW estimator. Third, RMST-based contrasts demonstrated more stable ordering and stronger separation among DTRs than contrasts based on survival probability, consistent with the fact that RMST aggregates information over time rather than relying on survival at a single time point. Consistent with the pattern observed for joint-model parameter estimation, the precision of DTR-specific estimates and pairwise contrasts improved for both methods as sample size increased (Supplementary Tables 1-2). The improvement was more pronounced for the IPTW estimator, reflecting the reduction in variability as the weighting scheme became more stable across treatment sequences.

For the secondary analysis, the sparse longitudinal measurement schedule compromised joint-model parameter estimation (Supplementary Table 3). Specifically, fixed effects in the longitudinal model demonstrated reduced coverage with moderate bias, and random-effect variance components showed substantial bias and severe undercoverage. With only four measurements per subject and truncation due to event or censoring, the data provided insufficient information to distinguish subject-specific variation from residual error. While most survival coefficients remained relatively stable, parameters depending more strongly on the longitudinal component, such as the baseline hazard scale and the current-value association parameter, showed increased bias compared with the main analysis. However, the primary conclusions of the joint model outperforming the IPTW estimator with regards to optimal regimen identification and DTR comparison, as well as the improvement pattern with increasing sample size, remained robust to reduced measurement frequency (Supplementary Tables 4-7). The efficiency advantage of the joint-model framework over the IPTW estimator persisted for both regimen-specific estimates and pairwise contrasts, indicating that these efficiency gains were not primarily driven by dense longitudinal measurements.

**7. Application to the AIPC trial**

We applied the proposed framework to data from a SMART for androgen-independent prostate cancer (AIPC) conducted at The University of Texas M.D. Anderson Cancer Center from December 1998 to January 2006.[51,52] The original trial was approved by the M.D. Anderson Cancer Center Institutional Review Board, and all participants provided written informed consent. The present study used de-identified data and was exempt from institutional review board review under U.S. federal regulations (45 CFR 46.104(d)(4)).

A total of 150 patients were randomized at enrollment to one of four chemotherapies: cyclophosphamide, vincristine, and dexamethasone (CVD); ketoconazole plus doxorubicin alternating with vinblastine plus estramustine (KA/VE); weekly paclitaxel, estramustine, and carboplatin (TEC); and paclitaxel, estramustine, and etoposide (TEE). Each treatment course lasted 8 weeks. At the end of each course, treatment was adapted based on a composite response indicator incorporating PSA reduction and disease status. Responders continued their current treatment, while nonresponders were rerandomized with equal probability to one of the remaining three treatments. This design embedded 12 DTRs as shown in Supplementary Figure 1.

In the original SMART design, patients received sequential treatments up to four courses, with follow-up terminating upon two consecutive responses or accumulating two non-responses, whichever occurred first, at any decision point. Substantial attrition occurred beyond the second course due to treatment-limiting toxicity and disease progression, with only 42 (28.0%) patients remaining evaluable at course 3 and 10 (6.6%) at course 4. To align with our proposed two-stage framework and mitigate bias from informative dropout, we restricted the analyses to the first two courses up to week 16, treating enrollment as the first stage and the response-dependent decision at week 8 as the second stage. PSA measurements at baseline, week 8, and week 16 served as the longitudinal outcome. Time to progression (TTP) was used as the primary endpoint as opposed to survival time since only 3 (2.0%) patients died within 16 weeks.

Of the 150 patients, 20 (13.3%) progressed before week 8 and contributed only first-stage information. An additional 11 (7.3%) patients discontinued protocol treatment prior to second-stage randomization due to toxicity, clinical decision, rapid progression, or withdrawal. These patients were treated as right-censored at week 8 under the assumption that censoring is independent of counterfactual TTP conditional on first-stage treatment and observed longitudinal history. A total of 374 PSA measurements were observed with 8 (2.1%) intermittently missing. Imputation was not performed as the likelihood-based estimation of the LMM accommodates intermittent missing data under a missing at random assumption.[53]

Within this 16-week two-stage SMART, first-stage randomization was well balanced across treatment arms (Supplementary Table 8). Among 119 patients who remained evaluable at week 8, 79 (66.4%) were classified as responders and continued their initial therapy, while only 40 (33.6%) were rerandomized to alternative treatments. Response rates varied substantially by first-stage treatment, with higher rates observed under KA/VE (66.7%) and TEC (65.8%), intermediate rates under TEE (51.3%), and notably lower rates under CVD (27.0%). Additionally, early progression prior to week 8 was also concentrated in the CVD arm (29.7%) and was rare under TEC (2.6%). These patterns indicated that the between-regimen separation was established before the second-stage decision, with patients initiating TEC or KA/VE entering the week-8 reassessment under more favorable risk profiles than those starting with CVD. Overall, 45 (30.0%) patients experienced progression, and 94 patients (62.7%) were administratively censored at week 16.

We fitted the joint model by maximum likelihood as described in Section 4. Given the limited sample size relative to the number of DTRs, we adopted parsimonious specifications that excluded baseline demographic and clinical covariates from both submodels. The LMM on the log-PSA scale included only cumulative time-on-treatment terms for TEE, TEC, and KA/VE as fixed effects (with CVD as the reference treatment), along with a random intercept. For the survival submodel, we assumed an exponential baseline hazard as an exploratory Weibull fit produced a shape parameter $\kappa$ close to 1 with less stable estimation; treatment effects were shared across stages for parsimony and numerical stability.

To implement the parametric g-formula, we incorporated a logistic response model within the joint likelihood since response in this trial was defined as a compound binary indicator. The response model with first-stage treatment indicators and the latent log-PSA value at week 8 as covariates was expressed as

$$\text{logit}\, P(R_i = 1 \mid V_{1i}, b_i) = \nu_0 + \nu_{\text{TEE}}\, I(V_{1i} = \text{TEE}) + \nu_{\text{TEC}}\, I(V_{1i} = \text{TEC}) + \nu_{\text{KA/VE}}\, I(V_{1i} = \text{KA/VE}) + \nu_{m(\tau)}\, m_i(\tau = 8)$$

Patients who progressed before week 8 or went off protocol at week 8 were excluded from the response likelihood and contributed only to first-stage likelihood components. DTR-specific progression-free survival probability (PFSP) and corresponding RMST at week 16, denoted as $S_{AIPC}(16)$ and $\Psi_{AIPC}(16)$, were estimated using both plug-in parametric g-formula under the fitted joint-model framework and the IPTW Kaplan–Meier estimator. Patients who experienced progression before week 8 or were censored at week 8 contributed to every DTR sharing their first-stage treatment. We then evaluated pairwise contrasts of DTR-specific estimates and implemented MCB with 1000 draws for the joint-model framework and 1000 bootstrap replicates for the IPTW estimator under $\varsigma = 0.05$.

As shown in Table 4, the fitted joint model revealed substantial between-patient heterogeneity in baseline PSA level ($\sigma_{b0}$=1.27, 95% CI: 1.11 to 1.45). The longitudinal submodel suggested insignificant differences in PSA trajectories among patients receiving TEE ($\beta_{\text{TEE}}$ = 0.41, 95% CI: −0.13 to 0.96), TEC ($\beta_{\text{TEC}}$ = 0.52, 95% CI: −0.15 to 1.20), or KA/VE ($\beta_{\text{KA/VE}}$ = 0.36, 95% CI: −0.18 to 0.91) relative to CVD. In contrast, the survival submodel indicated that TEC ($\text{HR}_{\text{TEC}}$=0.13, 95% CI: 0.04 to 0.50) and KA/VE ($\text{HR}_{\text{KA/VE}}$ = 0.32, 95% CI: 0.11 to 0.91) were associated with significantly reduced hazards of progression compared with CVD. Additionally, higher PSA levels were associated with increased hazards of progression ($\text{HR}_{m(t)}$ =1.28, 95% CI:

1.002 to 1.622). In the response model, lower latent PSA level at week 8 corresponded to a higher probability of response ($\mathrm{OR}_{m(\tau)}$ = 0.39, 95% CI: 0.24 to 0.63); while the treatment effects were estimated imprecisely with large ORs and extremely wide CIs, reflecting estimation uncertainty due to small sample size and limited number of non-responders.

According to Table 5, the joint-model and IPTW estimators identified the same top-ranked DTR, (TEC, TEC, KA/VE), and the same bottom-ranked regimen, (CVD, CVD, TEE), based on $\Psi_{AIPC}(16)$. These two approaches also showed general agreement in the rank of TEC-starting and CVD-starting DTRs. However, with only 40 nonresponders, DTR-specific estimates were largely driven by the first-stage treatment and the responder stratum, and the separation among DTRs sharing the same first-stage treatment was less pronounced (Figure 2A). Additionally, the middle of the DTR ranking differed substantially between the two methods and might be explained by the information usage. The joint-model framework incorporates information from all subjects through shared longitudinal, survival, and response components, while the IPTW estimator relies more heavily on a limited number of weighted observations within each treatment sequence. The joint-model framework also demonstrated modest gains in efficiency. For several DTRs with high $\Psi_{AIPC}(16)$ estimates, IPTW-based confidence intervals reached the upper bound of 16 weeks. After accounting for this ceiling effect, the joint-model framework still generated slightly narrower pointwise 95% confidence intervals for most DTR-specific RMST estimates.

Consistent with the simulation study, RMST provided more reliable discrimination among DTRs than survival probability in this trial. Based on $S_{AIPC}(16)$, both estimators identified the same top-ranked DTR but differed in the bottom-ranked DTR, and the ranking of DTRs demonstrated greater variability between estimators (Supplementary Table 9, Figure 2B). Notably, IPTW estimates of $S_{AIPC}(16)$ ranked two CVD-starting DTRs, (CVD, CVD, TEC) and (CVD,

CVD, KA/VE), above several non-CVD-starting DTRs, which was inconsistent with the poorer first-stage performance of CVD. This pattern likely reflected the limited observed events among those undergoing second-stage treatment switching. For example, no post-week-8 progression was observed among patients receiving $d$(CVD, TEC) or $d$(CVD, TEE) (Supplementary Table 8). The IPTW estimator upweights patients who follow these treatment sequences, so these event-free observations can inflate the estimated survival probabilities. This instability was also reflected in the MCB analyses based on $S_{AIPC}(16)$. The joint model retained all 12 regimens in the best set with limited discrimination, and the IPTW included two CVD-starting DTRs (Supplementary Table 9).

For the MCB results based on $\Psi_{AIPC}(16)$, all CVD-starting DTRs were excluded from the best set under both methods (Table 5). Regarding the size of the optimal set, the IPTW approach was more selective and retained seven DTRs, compared with nine under the joint-model framework. As described in Section 5.4, the joint-model framework employed a full covariance matrix obtained from a common fitted model, which induced dependence across DTR estimates. In the simulation study, the precision advantage of the joint model led to a smaller best set than the IPTW estimator. However, in the AIPC trial with limited second-stage rerandomized data, the joint model produced less separated estimates among second-stage strategies and led to more conservative multiplicity-adjusted comparisons.

To assess the robustness of the joint-model-based inference, we performed a series of diagnostic assessments to evaluate the specification of the longitudinal and survival submodels (Figure 3). For the longitudinal submodel, the observed and fitted PSA trajectories were closely aligned across all first-stage treatment groups; residual diagnostics indicated no major departures from mean specification, homoscedasticity, or normality; empirical Bayes estimates of the random intercepts showed substantial between-subject variability consistent with the model assumptions.

For the survival component, the Cox–Snell residual plot supported the exponential baseline hazard, with minor deviations likely due to limited events and high proportion of administrative censoring; calibration of the marginal joint-model-based survival curve against the overall Kaplan–Meier estimate demonstrated close agreement, indicating that the survival submodel provided a reasonable representation of the observed event-time distribution.

Our findings were consistent with the common findings in previous analyses of this trial, which identified TEC as the most favorable first-line therapy and CVD as comparatively inferior.[51,52] In agreement with Thall et al., both approaches identified TEC followed by KA/VE for nonresponders as the most favorable DTR and CVD followed by TEE as the least favorable based on $\Psi_{AIPC}(16)$.[51] However, based on expert-score outcomes over 32-week follow-up, Wang et al. favored CVD as the preferred salvage treatment after TEC.[52] Importantly, they also reported weak second-stage distinctions, with all 12 DTRs retained in the confidence set for the optimal rule according to the simultaneous confidence intervals they constructed. This lack of separation was consistent with our findings under MCB for $S_{AIPC}(16)$. Therefore, the difference in preferred salvage treatment likely arose from different estimands and follow-up horizons.

**8. Discussion**

We propose a joint longitudinal-survival framework for evaluating embedded DTRs in a two-stage SMART with repeated biomarker measurements and censored survival outcomes. This framework extends existing methods for survival data from SMARTs by exploiting the prognostic information captured by repeatedly measured biomarker data to inform estimation of DTR-specific survival outcomes, rather than using biomarker information only to define response. Through an extensive simulation study and an application to the AIPC trial,[51,52] we demonstrate that our framework provides unbiased estimates under correct model specification and substantial efficiency gains for DTR-specific survival inference.

Our joint-model framework advances existing approaches for DTR evaluation with survival outcomes through the following features: (i) incorporation of the longitudinal biomarker process into survival inference through a shared latent structure, while inheriting the standard advantages of joint models over the conventional time-dependent Cox model, including correction for measurement error, interpolation between intermittent longitudinal measurements, and proper handling of event-dependent dropout in the longitudinal process; (ii) specification of the treatment effects through a piecewise structure in both the longitudinal and survival submodels, which accommodates the sequential treatment assignment of SMARTs and can be adapted to alternative SMART designs by modifying the treatment indicators and piecewise structure without reconfiguring weighting schemes; (iii) flexible model specification that allows adjustment for baseline and post-decision-point covariates, as well as interactions between treatments and covariates to identify subgroups that benefit differentially from DTRs; (iv) derivation of explicit integral representations for the DTR-specific marginal survival function through the parametric g-formula in a single unified computation.

In our simulation study, the advantages of our joint-model framework manifested as uniformly smaller standard errors, higher optimal-regimen selection accuracy, and more discriminating MCB confidence sets. However, the efficiency gains require correct model specification of the longitudinal trajectory, the hazard function, the association structure, and the response model used in the g-formula. Therefore, rigorous diagnostic assessment of the fitted joint model is essential and should be a standard component of the analysis. In settings with strong concerns about model adequacy, the IPTW estimator may be preferable. Additionally, the use of doubly robust estimators, under which consistency is achieved if either the outcome model or the treatment-assignment model is correctly specified, could help preserve the efficiency advantages of the joint-model framework while providing a safeguard against model misspecification.[46]

The MCB results in the AIPC trial application highlight that reliable DTR discrimination requires adequate sample size, sufficient rerandomization at each decision point, and enough post-decision-point events. With only 40 rerandomized nonresponders and sparse second-stage treatment sequences in the AIPC trial data, neither the proposed joint model estimator nor the IPTW Kaplan–Meier estimator achieved meaningful separation among second-stage strategies, and most non-CVD regimens remained statistically indistinguishable from the estimated optimum. Additionally, as shown in our simulation study secondary analysis, the MCB procedure continued to identify the optimal DTR with high accuracy even when the number of longitudinal measurements was small relative to the number of random-effect parameters. However, the joint model had difficulty separating subject-specific variability from residual error and further compromised estimation of the latent trajectory in the survival submodel. These findings motivate future SMART designs with adequate second-stage rerandomization and frequent biomarker collection. Formal sample size determination warrants further investigation.

Our study has several limitations. SMARTs with extensive longitudinal biomarker and survival data remain rare, potentially due to the long follow-up required, the cost of administering multiple sequential treatments, and the need for well-defined intermediate tailoring variables. The AIPC trial applied in our analysis was designed as a hypothesis-generating study,[51] which restricts the power to draw confirmatory conclusions comparing 12 embedded DTRs and fully demonstrate the efficiency advantages of our joint-model framework. Beyond data limitations, the joint-model framework is computationally intensive. In our simulation, mean optimization times were around 11, 22, and 38 minutes at N = 300, 600, and 1,200, respectively, and this burden increased further when longitudinal measurements were sparse, as the likelihood surface became less regular and the optimizer required more iterations to converge; the MCB and AESE inference required around 5 hours per replication at N=300 due to repeated computations of regimen-specific values across multivariate normal draws and bootstrap resamples. For the AIPC trial application, however, the entire procedure completed within an hour. Furthermore, the causal interpretation of second-stage treatment continuation effects under the tailoring-variable SMART design warrants caution. Responders continue their first-stage treatment rather than being rerandomized, so the contrasts between continuation and switching are confounded by response status. The corresponding parameters should be interpreted as pathway-specific associations rather than pure second-stage treatment causal effects. Importantly, this limitation does not undermine the causal interpretation of the regimen-specific survival estimands obtained from the parametric g-formula, which remain well-defined marginal counterfactual outcomes because they integrate over the regimen-specific response distribution. To enable unconfounded estimation of second-stage treatment effects for both responders and nonresponders, future studies should consider SMART designs with full representation of all second-stage treatments through randomization.

The framework developed in this article can be extended in several future directions. First, the current-value association parameterization can be replaced or supplemented by alternative functional forms, such as the rate of change, cumulative exposure, or shared random-effect parameterizations, which may better capture the biological mechanism linking the biomarker to the event in specific clinical contexts.[38] Second, while we adopt a parametric baseline hazard for computational simplicity, flexible alternatives such as piecewise constant or spline-based baseline hazards can be explored to reduce sensitivity to distributional assumptions. Third, the framework may accommodate multiple longitudinal outcomes for binary, continuous and ordinal biomarkers by assuming a joint multivariate normal distribution across all outcome-specific random effects.[54] Fourth, dynamic prediction of patient-specific survival outcomes conditional on their biomarker history is a natural extension that can inform real-time treatment adaptation within a SMART, as the fitted joint model enables individualized survival comparisons across candidate second-stage treatments by combining a patient's random-effect posterior with population-level treatment coefficients estimated from previous patients. Finally, the integration of this joint-model framework with Bayesian methods and informative prior information may offer advantages for quantifying uncertainty in settings with small samples and complex model structures.

In conclusion, the proposed joint longitudinal-survival framework provides a comprehensive characterization of how treatment sequences influence both biomarker evolution and clinical outcomes, enables valid and efficient causal inference for DTR-specific survival, and supports formal multiplicity-adjusted comparisons among embedded DTRs. Despite the natural suitability of SMART designs for oncology clinical trials with survival endpoints and longitudinal biomarker monitoring, their adoption remains limited in practice. We hope our proposed framework will lower methodological barriers to adopting SMART designs, facilitate the construction of evidence-based adaptive treatment strategies, and motivate broader use of sequential randomization in oncology clinical trials.

**Acknowledgments**
This work made use of the High Performance Computing Resource in the Core Facility for Advanced Research Computing at Case Western Reserve University. We thank Dr. Lu Wang from the University of Michigan for providing access to the androgen-independent prostate cancer trial data.

**Funding Statement**
This research received no specific grant from any funding agency in the public, commercial, or not-for-profit sectors.

**Conflict of Interest Statement**
The authors have declared no conflict of interest.

**Data Availability Statement**
The androgen-independent prostate cancer trial data analyzed in this study are available from Dr. Lu Wang, University of Michigan, upon reasonable request. Simulated data, simulation code and analysis code are available from the corresponding author upon request.

**Tables**

Table 1. Maximum likelihood estimation of joint-model parameters with a maximum of 25 longitudinal measurements per subject across 1000 replications.

| Parameter | True value | Rel% | MCSE | AESE | RMSE | Cov% | Rel% | MCSE | AESE | RMSE | Cov% | Rel% | MCSE | AESE | RMSE | Cov% |
|---|---|---|---|---|---|---|---|---|---|---|---|---|---|---|---|---|
| | | N=300 | | | | | N=600 | | | | | N=1200 | | | | |
| Longitudinal Submodel | | | | | | | | | | | | | | | | |
| $\beta_0$ | 3.50 | −0.0083 | 0.0485 | 0.0484 | 0.0485 | 95.6 | −0.0062 | 0.0337 | 0.0342 | 0.0336 | 95.4 | 0.0048 | 0.0250 | 0.0242 | 0.0250 | 94.9 |
| $\beta_{X_{01}}$ | 0.50 | −0.0906 | 0.0604 | 0.0596 | 0.0603 | 94.3 | 0.1867 | 0.0403 | 0.0421 | 0.0403 | 96.3 | 0.0042 | 0.0303 | 0.0298 | 0.0302 | 95.2 |
| $\beta_{X_{02}}$ | 0.70 | 0.2023 | 0.0304 | 0.0293 | 0.0304 | 94.6 | 0.0607 | 0.0212 | 0.0207 | 0.0212 | 94.6 | 0.0890 | 0.0150 | 0.0147 | 0.0150 | 94.4 |
| $\beta_t$ | −0.50 | −0.3498 | 0.0631 | 0.0602 | 0.0631 | 93.8 | 0.3657 | 0.0444 | 0.0428 | 0.0444 | 93.3 | 0.6173 | 0.0299 | 0.0302 | 0.0300 | 94.6 |
| $\beta_A$ | −0.80 | 0.3440 | 0.0754 | 0.0721 | 0.0754 | 93.9 | −0.0860 | 0.0521 | 0.0512 | 0.0521 | 94.9 | −0.5071 | 0.0359 | 0.0361 | 0.0361 | 94.8 |
| $\beta_B$ | −0.60 | 0.6001 | 0.0767 | 0.0744 | 0.0768 | 94.8 | −0.0607 | 0.0556 | 0.0528 | 0.0556 | 93.8 | −0.6041 | 0.0376 | 0.0373 | 0.0378 | 94.5 |
| $\beta_C$ | −0.70 | 0.3380 | 0.0781 | 0.0760 | 0.0781 | 94.3 | −0.2441 | 0.0555 | 0.0541 | 0.0555 | 94.4 | −0.4624 | 0.0380 | 0.0382 | 0.0381 | 94.5 |
| $\sigma_{b_0}$ | 0.50 | −0.4063 | 0.0242 | 0.0249 | 0.0242 | 95.2 | −0.4172 | 0.0176 | 0.0176 | 0.0177 | 94.3 | −0.0887 | 0.0123 | 0.0125 | 0.0123 | 94.6 |
| $\sigma_{b_1}$ | 0.20 | −2.2449 | 0.0215 | 0.0211 | 0.0220 | 95.1 | −0.7400 | 0.0148 | 0.0148 | 0.0149 | 94.5 | −0.5730 | 0.0103 | 0.0104 | 0.0104 | 95.5 |
| $\rho$ | −0.30 | −1.6028 | 0.1026 | 0.1007 | 0.1027 | 95.6 | 0.7233 | 0.0710 | 0.0705 | 0.0710 | 95.3 | 0.2855 | 0.0481 | 0.0495 | 0.0481 | 96.2 |
| $\sigma_\varepsilon$ | 0.50 | −0.0105 | 0.0058 | 0.0057 | 0.0058 | 94.4 | 0.0029 | 0.0040 | 0.0040 | 0.0040 | 95.2 | −0.0070 | 0.0029 | 0.0029 | 0.0029 | 94.6 |
| Survival Submodel | | | | | | | | | | | | | | | | |
| $\lambda_0$ | 0.15 | 11.0959 | 0.0910 | 0.0814 | 0.0925 | 94.8 | 5.6914 | 0.0543 | 0.0541 | 0.0549 | 95.7 | 2.1973 | 0.0378 | 0.0368 | 0.0379 | 95.1 |
| $\kappa$ | 2.60 | 1.5288 | 0.1991 | 0.1948 | 0.2029 | 93.7 | 0.5806 | 0.1323 | 0.1357 | 0.1331 | 95.9 | 0.3123 | 0.0987 | 0.0955 | 0.0989 | 93.8 |
| $\gamma_{X_{01}}$ | 0.40 | 1.1596 | 0.1901 | 0.1836 | 0.1900 | 94.6 | 1.9394 | 0.1270 | 0.1282 | 0.1272 | 95.9 | −0.3367 | 0.0922 | 0.0900 | 0.0921 | 94.4 |
| $\gamma_{X_{02}}$ | 0.20 | 1.7263 | 0.1426 | 0.1418 | 0.1426 | 94.8 | 2.9770 | 0.0990 | 0.0985 | 0.0991 | 95.2 | 0.1312 | 0.0697 | 0.0692 | 0.0697 | 94.6 |
| $\gamma_A$ | −0.50 | −3.6058 | 0.2519 | 0.2505 | 0.2524 | 93.9 | −1.6484 | 0.1751 | 0.1758 | 0.1753 | 95.6 | −0.3819 | 0.1269 | 0.1237 | 0.1268 | 94.6 |
| $\gamma_{AA}$ | −1.50 | −3.8058 | 0.4933 | 0.4707 | 0.4964 | 95.5 | −1.8816 | 0.3292 | 0.3241 | 0.3302 | 95.3 | −0.8784 | 0.2286 | 0.2267 | 0.2289 | 94.5 |
| $\gamma_{BB}$ | −1.40 | −2.8648 | 0.4804 | 0.4605 | 0.4818 | 95.5 | −1.6079 | 0.3120 | 0.3155 | 0.3127 | 95.6 | −1.2804 | 0.2208 | 0.2218 | 0.2214 | 95.2 |
| $\gamma_{AC}$ | −1.00 | −3.2733 | 0.4124 | 0.4015 | 0.4135 | 95.7 | −1.1896 | 0.2791 | 0.2790 | 0.2792 | 95.0 | 0.2090 | 0.1926 | 0.1954 | 0.1925 | 95.2 |
| $\gamma_{BC}$ | −0.90 | −2.0466 | 0.3778 | 0.3687 | 0.3781 | 94.8 | −1.8583 | 0.2531 | 0.2568 | 0.2536 | 95.8 | −0.2457 | 0.1796 | 0.1798 | 0.1796 | 94.9 |
| $\alpha$ | 0.20 | 2.2097 | 0.1683 | 0.1679 | 0.1683 | 94.7 | −0.5966 | 0.1149 | 0.1168 | 0.1149 | 94.2 | 1.4158 | 0.0817 | 0.0819 | 0.0817 | 94.9 |

Note: Rel% = relative bias in %; MCSE = empirical standard deviation of point estimates across replications; AESE = the average of estimated standard errors from the inverse observed information matrix; RMSE = root mean squared error; Cov% = coverage probability of the nominal 95% Wald confidence interval based on AESE.

Table 2. Regimen-specific estimates, optimal regimen selection and MCB best-set inclusion for N = 300 with a maximum of 25 longitudinal measurements per subject across 1000 replications.

| Estimand | DTR | True value | Joint-model framework | | | | | | | IPTW Kaplan–Meier | | | | | | | RE |
|---|---|---|---|---|---|---|---|---|---|---|---|---|---|---|---|---|---|
| | | | Rel% | MCSE | AESE | RMSE | Cov% | Point% | MCB% | Rel% | MCSE | AESE | RMSE | Cov% | Point% | MCB% | |
| $\Psi(16)$ | (A,A,C) | 13.3525 | 0.1396 | 0.2832 | 0.2773 | 0.2837 | 94.3 | 99.5 | 100.0 | 0.0678 | 0.3727 | 0.3810 | 0.3727 | 95.2 | 71.0 | 99.6 | 0.58 |
| | (A,A,D) | 13.1311 | 0.1056 | 0.3016 | 0.2948 | 0.3018 | 93.9 | 0.0 | 9.0 | 0.2227 | 0.3727 | 0.3791 | 0.3737 | 94.7 | 24.2 | 94.6 | 0.66 |
| | (B,B,C) | 12.4664 | −0.0378 | 0.3159 | 0.2953 | 0.3158 | 92.3 | 0.5 | 22.0 | −0.0709 | 0.4081 | 0.4135 | 0.4080 | 94.5 | 4.0 | 63.5 | 0.60 |
| | (B,B,D) | 12.1957 | −0.0701 | 0.3270 | 0.3101 | 0.3270 | 92.1 | 0.0 | 2.6 | −0.0782 | 0.3951 | 0.4041 | 0.3951 | 94.0 | 0.8 | 40.6 | 0.68 |
| $\Psi(24)$ | (A,A,C) | 17.5462 | 0.1854 | 0.5835 | 0.5675 | 0.5841 | 94.7 | 99.3 | 100.0 | 0.0602 | 0.7266 | 0.7388 | 0.7263 | 95.1 | 94.3 | 100.0 | 0.64 |
| | (A,A,D) | 16.2729 | 0.1567 | 0.5934 | 0.5764 | 0.5936 | 94.3 | 0.0 | 7.7 | 0.2671 | 0.6883 | 0.6856 | 0.6894 | 95.2 | 4.1 | 64.6 | 0.74 |
| | (B,B,C) | 15.4195 | −0.1068 | 0.6090 | 0.5629 | 0.6089 | 92.5 | 0.7 | 17.0 | −0.1038 | 0.7490 | 0.7508 | 0.7487 | 94.7 | 1.6 | 47.8 | 0.66 |
| | (B,B,D) | 14.1533 | −0.1042 | 0.5613 | 0.5398 | 0.5612 | 93.5 | 0.0 | 0.1 | −0.1676 | 0.6414 | 0.6518 | 0.6416 | 94.3 | 0.0 | 6.6 | 0.77 |
| $S(16)$ | (A,A,C) | 0.5994 | 0.3844 | 0.0360 | 0.0352 | 0.0360 | 94.5 | 99.4 | 100.0 | −0.1012 | 0.0516 | 0.0532 | 0.0515 | 94.6 | 85.3 | 99.9 | 0.49 |
| | (A,A,D) | 0.5227 | 0.2343 | 0.0406 | 0.0396 | 0.0406 | 94.3 | 0.0 | 8.4 | 0.5730 | 0.0548 | 0.0553 | 0.0548 | 94.6 | 11.9 | 82.1 | 0.55 |
| | (B,B,C) | 0.4613 | −0.2915 | 0.0374 | 0.0348 | 0.0374 | 91.9 | 0.6 | 16.8 | −0.0808 | 0.0555 | 0.0558 | 0.0555 | 94.5 | 2.8 | 60.0 | 0.45 |
| | (B,B,D) | 0.3747 | −0.4850 | 0.0385 | 0.0375 | 0.0385 | 93.7 | 0.0 | 0.3 | −0.4105 | 0.0539 | 0.0547 | 0.0539 | 94.4 | 0.0 | 15.6 | 0.51 |
| $S(24)$ | (A,A,C) | 0.4659 | 0.1952 | 0.0536 | 0.0515 | 0.0536 | 94.3 | 99.0 | 100.0 | 0.0569 | 0.0574 | 0.0566 | 0.0574 | 94.6 | 98.0 | 100.0 | 0.87 |
| | (A,A,D) | 0.2907 | 0.6118 | 0.0389 | 0.0379 | 0.0389 | 95.1 | 0.0 | 7.7 | 0.6674 | 0.0503 | 0.0508 | 0.0503 | 95.5 | 0.3 | 25.9 | 0.60 |
| | (B,B,C) | 0.3003 | −0.4066 | 0.0512 | 0.0476 | 0.0512 | 93.3 | 1.0 | 32.8 | −0.4472 | 0.0542 | 0.0535 | 0.0542 | 94.1 | 1.7 | 42.7 | 0.89 |
| | (B,B,D) | 0.1556 | 0.1773 | 0.0297 | 0.0294 | 0.0297 | 94.3 | 0.0 | 0.0 | −1.4531 | 0.0390 | 0.0395 | 0.0391 | 94.2 | 0.0 | 0.4 | 0.58 |

Note: $S(t)$ = survival function at time $t$ = 16, 24 weeks; $\Psi(t^*)$ = restricted mean survival time at time horizon $t^*$ = 16, 24 weeks; DTR = dynamic treatment regimen; Rel% = relative bias in %; MCSE = empirical standard deviation of point estimates across replications; AESE = the average of estimated standard errors from the multivariate normal propagation for joint-model framework and the bootstrap replicates for IPTW estimator; RMSE = root mean squared error; Cov% = coverage probability of the nominal 95% Wald confidence interval based on AESE; Point % = percentage of simulations selecting each DTR as optimal based on point-estimate; MCB% = percentage of simulations including each DTR in the MCB best set; RE = relative efficiency (MCSE-based variance ratio of joint-model framework to IPTW Kaplan-Meier).

Table 3. Pairwise regimen contrasts for N = 300 with a maximum of 25 longitudinal measurements per subject across 1000 replications.

| Estimand | Contrast | True value | Joint-model framework | | | | | IPTW Kaplan–Meier | | | | | RE |
|---|---|---|---|---|---|---|---|---|---|---|---|---|---|
| | | | Rel% | MCSE | AESE | RMSE | Cov% | Rel% | MCSE | AESE | RMSE | Cov% | |
| Ψ(16) | (A,A,C)−(A,A,D) | 0.2214 | 2.1549 | 0.0744 | 0.0726 | 0.0745 | 93.9 | −9.1199 | 0.3043 | 0.3450 | 0.3048 | 97.1 | 0.06 |
| | (A,A,C)−(B,B,C) | 0.8861 | 2.6358 | 0.3345 | 0.3317 | 0.3352 | 93.5 | 2.0187 | 0.5469 | 0.5626 | 0.5469 | 95.5 | 0.37 |
| | (A,A,C)−(B,B,D) | 1.1569 | 2.3502 | 0.3611 | 0.3600 | 0.3619 | 93.4 | 1.6066 | 0.5444 | 0.5552 | 0.5444 | 95.3 | 0.44 |
| | (A,A,D)−(B,B,C) | 0.6647 | 2.7959 | 0.3627 | 0.3557 | 0.3630 | 93.1 | 5.7286 | 0.5473 | 0.5614 | 0.5484 | 95.9 | 0.44 |
| | (A,A,D)−(B,B,D) | 0.9355 | 2.3964 | 0.3848 | 0.3799 | 0.3853 | 93.0 | 4.1452 | 0.5468 | 0.5538 | 0.5479 | 94.8 | 0.50 |
| | (B,B,C)−(B,B,D) | 0.2707 | 1.4153 | 0.0892 | 0.0879 | 0.0892 | 95.1 | 0.2575 | 0.3382 | 0.4034 | 0.3380 | 97.6 | 0.07 |
| Ψ(24) | (A,A,C)−(A,A,D) | 1.2732 | 0.5526 | 0.3814 | 0.3691 | 0.3813 | 93.9 | −2.5851 | 0.7067 | 0.7322 | 0.7071 | 94.9 | 0.29 |
| | (A,A,C)−(B,B,C) | 2.1267 | 2.3037 | 0.7351 | 0.7272 | 0.7364 | 95.6 | 1.2488 | 1.0307 | 1.0534 | 1.0305 | 94.8 | 0.51 |
| | (A,A,C)−(B,B,D) | 3.3929 | 1.3933 | 0.7560 | 0.7542 | 0.7571 | 94.4 | 1.0103 | 0.9666 | 0.9846 | 0.9667 | 95.1 | 0.61 |
| | (A,A,D)−(B,B,C) | 0.8535 | 4.9162 | 0.7940 | 0.7631 | 0.7947 | 93.3 | 6.9683 | 1.0119 | 1.0172 | 1.0131 | 95.8 | 0.62 |
| | (A,A,D)−(B,B,D) | 2.1196 | 1.8983 | 0.7801 | 0.7631 | 0.7807 | 93.5 | 3.1701 | 0.9517 | 0.9453 | 0.9536 | 93.9 | 0.67 |
| | (B,B,C)−(B,B,D) | 1.2661 | −0.1360 | 0.3955 | 0.3835 | 0.3953 | 94.3 | 0.6099 | 0.7081 | 0.7574 | 0.7078 | 95.9 | 0.31 |
| $S(16)$ | (A,A,C)−(A,A,D) | 0.0767 | 1.4068 | 0.0243 | 0.0236 | 0.0243 | 94.3 | −4.6931 | 0.0618 | 0.0643 | 0.0619 | 95.3 | 0.15 |
| | (A,A,C)−(B,B,C) | 0.1382 | 2.6410 | 0.0478 | 0.0474 | 0.0479 | 95.3 | −0.1690 | 0.0747 | 0.0771 | 0.0747 | 95.1 | 0.41 |
| | (A,A,C)−(B,B,D) | 0.2247 | 1.8342 | 0.0521 | 0.0520 | 0.0522 | 95.1 | 0.4146 | 0.0743 | 0.0762 | 0.0743 | 94.9 | 0.49 |
| | (A,A,D)−(B,B,C) | 0.0614 | 4.1828 | 0.0537 | 0.0518 | 0.0537 | 93.5 | 5.4825 | 0.0773 | 0.0785 | 0.0774 | 95.8 | 0.48 |
| | (A,A,D)−(B,B,D) | 0.1480 | 2.0559 | 0.0554 | 0.0542 | 0.0554 | 93.8 | 3.0632 | 0.0776 | 0.0777 | 0.0777 | 94.3 | 0.51 |
| | (B,B,C)−(B,B,D) | 0.0866 | 0.5464 | 0.0273 | 0.0267 | 0.0273 | 94.4 | 1.3462 | 0.0669 | 0.0681 | 0.0669 | 94.6 | 0.17 |
| $S(24)$ | (A,A,C)−(A,A,D) | 0.1752 | −0.4960 | 0.0503 | 0.0484 | 0.0503 | 93.2 | −0.9558 | 0.0645 | 0.0637 | 0.0645 | 94.4 | 0.61 |
| | (A,A,C)−(B,B,C) | 0.1657 | 1.2859 | 0.0705 | 0.0689 | 0.0705 | 95.0 | 0.9705 | 0.0775 | 0.0779 | 0.0775 | 95.5 | 0.83 |
| | (A,A,C)−(B,B,D) | 0.3103 | 0.2042 | 0.0613 | 0.0596 | 0.0612 | 94.6 | 0.8139 | 0.0689 | 0.0692 | 0.0689 | 95.5 | 0.79 |
| | (A,A,D)−(B,B,C) | −0.0096 | −31.3739 | 0.0639 | 0.0604 | 0.0640 | 93.3 | −34.3370 | 0.0755 | 0.0738 | 0.0755 | 93.6 | 0.72 |
| | (A,A,D)−(B,B,D) | 0.1351 | 1.1121 | 0.0464 | 0.0450 | 0.0464 | 94.7 | 3.1090 | 0.0642 | 0.0645 | 0.0643 | 94.8 | 0.52 |
| | (B,B,C)−(B,B,D) | 0.1447 | −1.0346 | 0.0467 | 0.0446 | 0.0467 | 93.6 | 0.6345 | 0.0556 | 0.0564 | 0.0555 | 94.7 | 0.71 |

Note: $S(t)$ = survival function at time $t$ = 16, 24 weeks; $\Psi(t^*)$ = restricted mean survival time at time horizon $t^*$ = 16, 24 weeks; Rel% = relative bias in %; MCSE = empirical standard deviation of point estimates across replications; AESE = the average of estimated standard errors from the multivariate normal propagation for joint-model framework and the bootstrap replicates for IPTW estimator; Cov% = coverage probability of the nominal 95% Wald confidence interval based on AESE; RE = relative efficiency (MCSE-based variance ratio of joint-model framework to IPTW Kaplan-Meier).

Table 4. Maximum likelihood estimates of the fitted joint model parameters for the androgen-independent prostate cancer trial.

| Parameter | Estimate | SE | 95% CI | Scale | HR/OR | HR/OR 95% CI |
|---|---|---|---|---|---|---|
| Longitudinal Submodel | | | | | | |
| $\beta_0$ | 3.3036 | 0.1028 | [3.1022, 3.5050] | -- | -- | -- |
| $\beta_{\mathrm{TEE}}$ | 0.4109 | 0.2776 | [−0.1333, 0.9551] | -- | -- | -- |
| $\beta_{\mathrm{TEC}}$ | 0.5249 | 0.3429 | [−0.1472, 1.1970] | -- | -- | -- |
| $\beta_{\mathrm{KA/VE}}$ | 0.3647 | 0.2789 | [−0.1820, 0.9114] | -- | -- | -- |
| $\sigma_{b_0}$ | 1.2687 | 0.0882 | [1.1072, 1.4539] | -- | -- | -- |
| $\sigma_{\varepsilon}$ | 0.9273 | 0.0423 | [0.8480, 1.0141] | -- | -- | -- |
| Survival Submodel | | | | | | |
| $\lambda_0$ | 0.0135 | 0.0076 | [0.0045, 0.0407] | -- | -- | -- |
| $\gamma_{\mathrm{TEE}}$ | −0.9545 | 0.5189 | [−1.9715, 0.0626] | HR | 0.39 | [0.14, 1.07] |
| $\gamma_{\mathrm{TEC}}$ | −2.0189 | 0.6731 | [−3.3382, −0.6996] | HR | 0.13 | [0.04, 0.50] |
| $\gamma_{\mathrm{KA/VE}}$ | −1.1489 | 0.5390 | [−2.2052, −0.0926] | HR | 0.32 | [0.11, 0.91] |
| $\alpha$ | 0.2429 | 0.1229 | [0.0019, 0.4838] | HR | 1.28 | [1.002, 1.622] |
| Logistic Response Submodel | | | | | | |
| $\nu_0$ | 3.0039 | 0.9918 | [1.0599, 4.9478] | OR | 20.16 | [2.89, 140.87] |
| $\nu_{\mathrm{TEE}}$ | 3.9727 | 2.2699 | [−0.4763, 8.4216] | OR | 53.13 | [0.62, 4544.17] |
| $\nu_{\mathrm{TEC}}$ | 4.9919 | 2.7498 | [−0.3976, 10.3814] | OR | 147.22 | [0.67, 32254.09] |
| $\nu_{\mathrm{KA/VE}}$ | 4.5713 | 2.3021 | [0.0592, 9.0834] | OR | 96.67 | [1.06, 8807.86] |
| $\nu_{\mathrm{m}(\tau)}$ | −0.9396 | 0.2437 | [−1.4172, −0.4620] | OR | 0.39 | [0.24, 0.63] |

Note: HR = hazard ratio; OR = odds ratio; 95% confidence intervals for strictly positive parameters ($\sigma_{b_0}$, $\sigma_{\varepsilon}$, $\lambda_0$) are constructed on the log scale and back transformed to ensure positivity, all other confidence intervals are standard Wald intervals.

Table 5. Regimen-specific progression-free restricted mean survival time estimates at week 16 and MCB best-set inclusion for the androgen-independent prostate cancer trial

| Joint-model framework | | | | IPTW Kaplan–Meier | | | |
|---|---|---|---|---|---|---|---|
| DTR | $\Psi_{AIPC}(16)$(95% CI) | $\mathbb{M}$ | In best set | DTR | $\Psi_{AIPC}(16)$(95% CI) | $\mathbb{M}$ | In best set |
| (TEC,TEC,KA/VE) | 15.0894 (14.1782, 15.6137) | 2.6776 | Yes | (TEC,TEC,KA/VE) | 15.6743 (15.0514, 16.0000) | 3.8336 | Yes |
| (TEC,TEC,TEE) | 15.0318 (14.1473, 15.5703) | 2.1121 | Yes | (TEC,TEC,CVD) | 15.4402 (14.7188, 16.0000) | 1.1143 | Yes |
| (TEC,TEC,CVD) | 14.6011 (13.6617, 15.4599) | 0.6251 | Yes | (TEE,TEE,CVD) | 14.3062 (13.0315, 15.3682) | 0.3985 | Yes |
| (KA/VE,KA/VE,TEC) | 14.5926 (13.0882, 15.3012) | 1.3702 | Yes | (KA/VE,KA/VE,CVD) | 14.2833 (12.9916, 15.3895) | 0.3672 | Yes |
| (KA/VE,KA/VE,TEE) | 14.4874 (13.0243, 15.2037) | 1.2211 | Yes | (KA/VE,KA/VE,TEC) | 14.1671 (12.8535, 15.3457) | 0.2640 | Yes |
| (TEE,TEE,TEC) | 14.3532 (12.6341, 15.1662) | 1.0671 | Yes | (TEC,TEC,TEE) | 14.1035 (12.7254, 15.6339) | 0.1569 | Yes |
| (KA/VE,KA/VE,CVD) | 14.2755 (12.9388, 15.0439) | 0.7180 | Yes | (TEE,TEE,TEC) | 13.8981 (12.3572, 15.1299) | −0.0009 | No |
| (TEE,TEE,KA/VE) | 14.2157 (12.5615, 15.0362) | 0.8937 | Yes | (KA/VE,KA/VE,TEE) | 13.8607 (12.6772, 14.9719) | −0.2964 | No |
| (TEE,TEE,CVD) | 13.7962 (12.3400, 14.7448) | 0.1439 | Yes | (TEE,TEE,KA/VE) | 13.7810 (12.1507, 15.2542) | 0.0680 | Yes |
| (CVD,CVD,TEC) | 12.9557 (11.0368, 14.0625) | −0.3482 | No | (CVD,CVD,KA/VE) | 13.1767 (11.2616, 14.5578) | −0.4137 | No |
| (CVD,CVD,KA/VE) | 12.8785 (10.9184, 14.0184) | −0.4098 | No | (CVD,CVD,TEC) | 12.8073 (10.6545, 14.3345) | −0.5895 | No |
| (CVD,CVD,TEE) | 12.8506 (10.9160, 14.0188) | −0.4145 | No | (CVD,CVD,TEE) | 12.2495 (10.0142, 14.0850) | −0.8579 | No |

Note: DTR = dynamic treatment regimen; $\Psi_{AIPC}(t^*)$ = progression-free restricted mean survival time at week 16; MCB = multiple comparisons with the best; $\mathbb{M}$ = multiplicity-adjusted MCB inclusion margin (positive = included, negative = excluded). MCB is implemented at significance level $\varsigma = 0.05$.

**Figures**

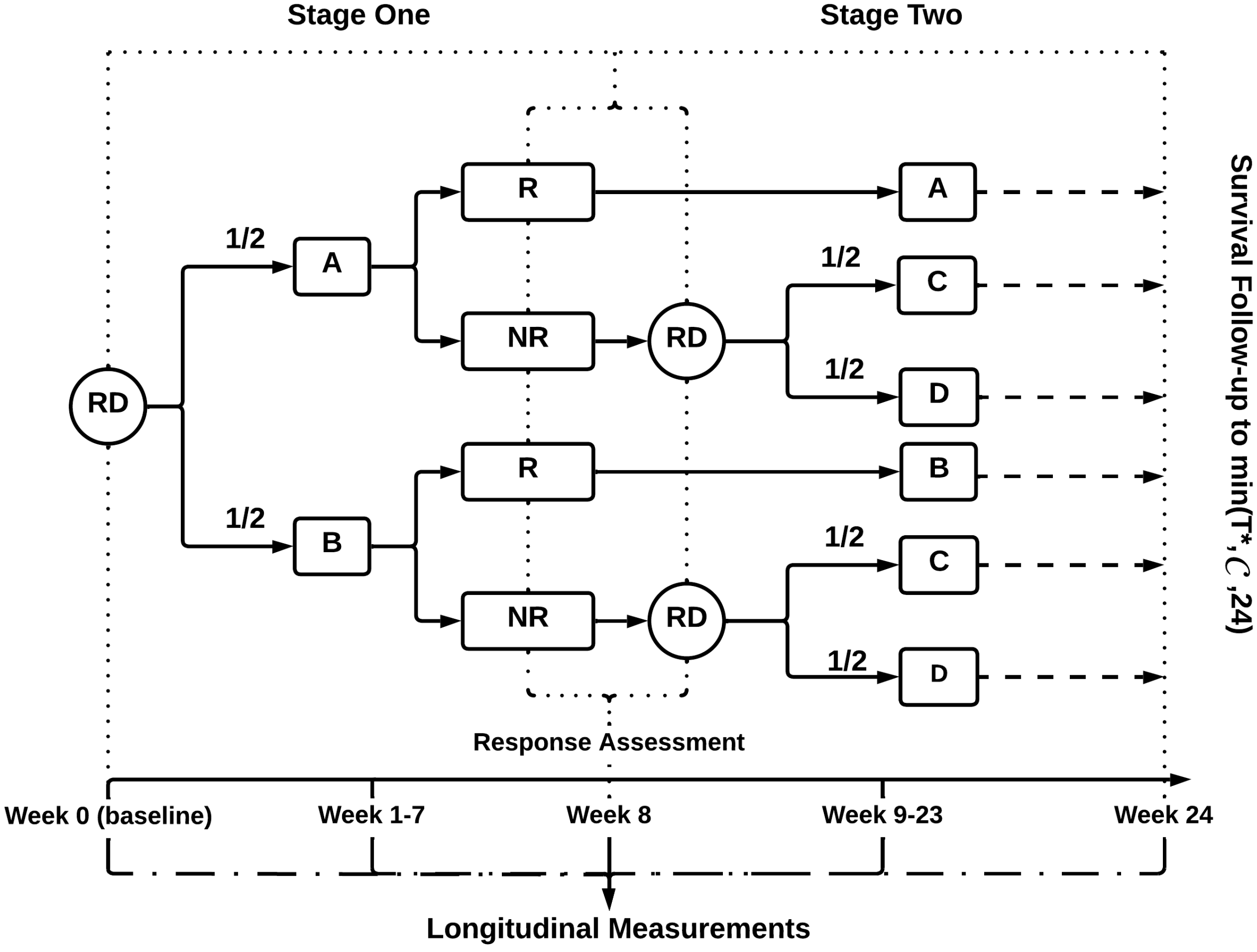

Figure 1. Schematic of the two-stage tailoring-variable sequential multiple assignment randomized trial design, longitudinal biomarker measurement schedule, and survival follow-up through administrative censoring. A, B, C and D represent the four available treatments; RD indicates a decision point for randomization; R and NR represent response and nonresponse; the expressions along the lines indicate treatment randomization probability; $T^*$ denotes the true event time and $\mathcal{C}$ represents the censoring time.

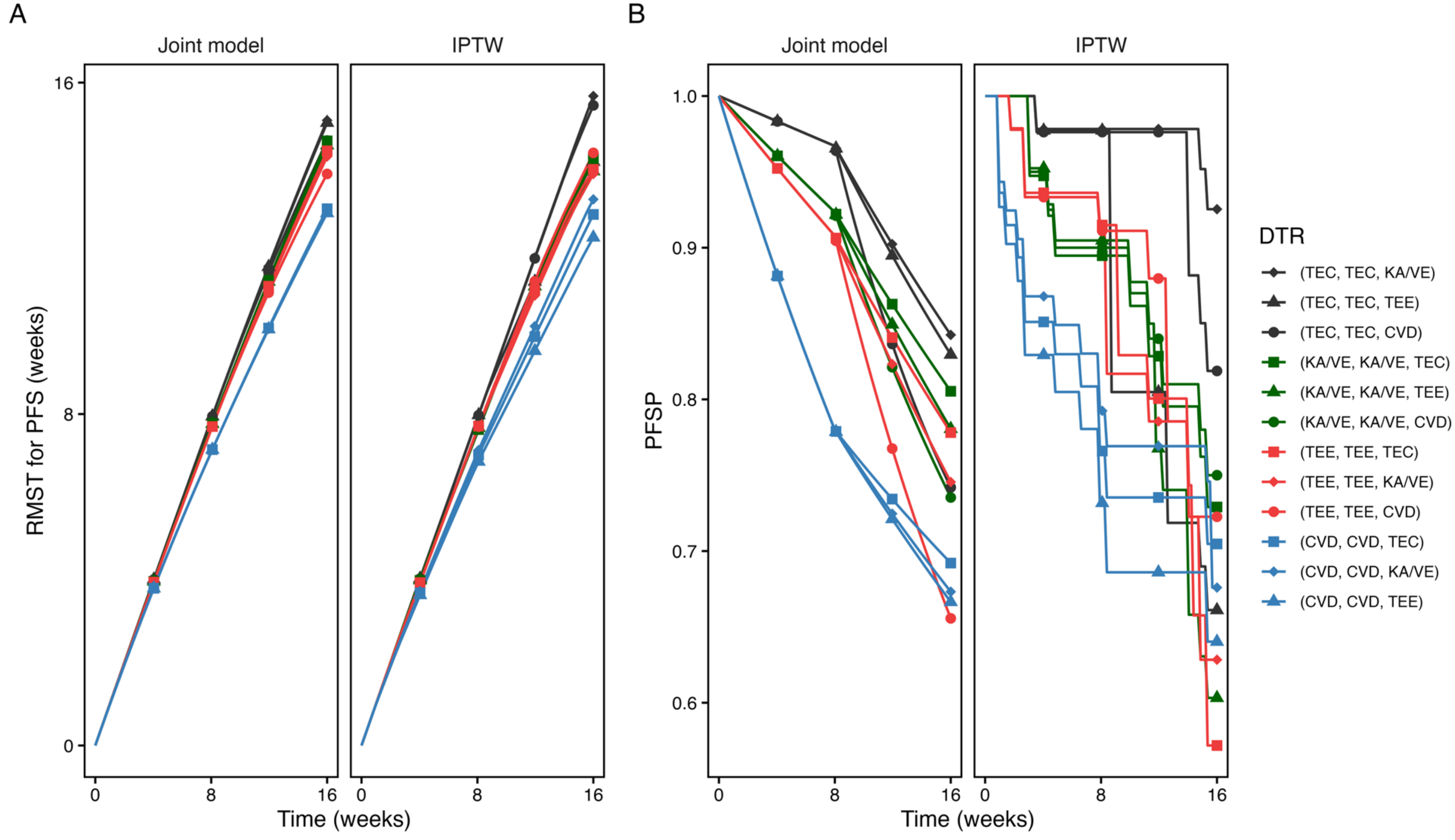

Figure 2. Dynamic treatment regimen (DTR)-specific (A) restricted mean survival times (RMST) and (B) progression-free survival probabilities (PFSP) from the baseline to week 16 for the androgen-independent prostate cancer trial, estimated by the joint-model framework and the inverse probability of treatment weighted (IPTW) Kaplan–Meier estimator.

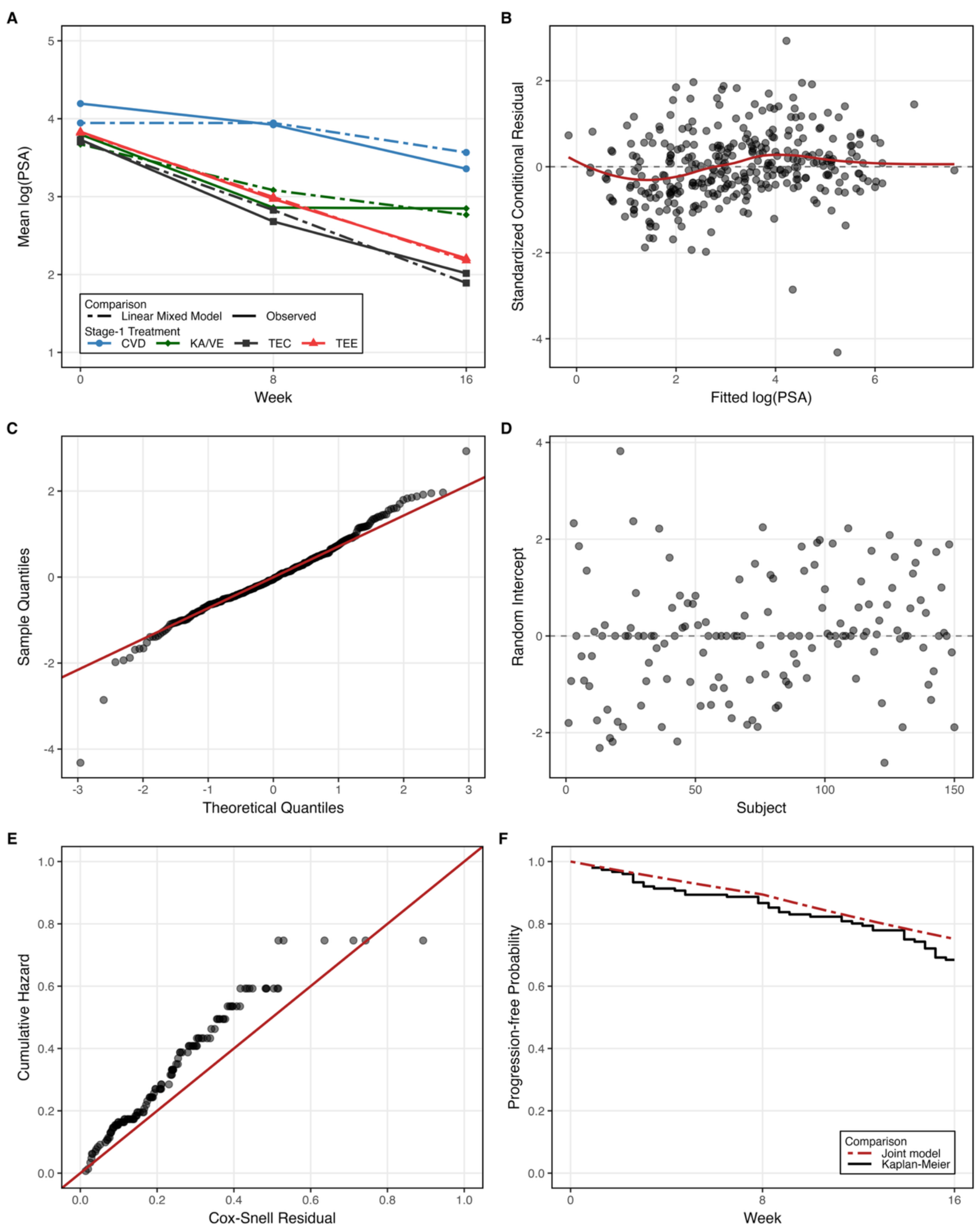

Figure 3. Diagnostic plots for the fitted joint model using the androgen-independent prostate cancer trial data: (A) observed and linear mixed model fitted mean log(PSA) trajectories by first-stage treatment; (B) standardized conditional residuals versus fitted values; (C) normal Q–Q plot of standardized conditional residuals; (D) empirical Bayes estimates of the subject-specific random intercepts from the fitted joint model; (E) Cox–Snell residual plot for the survival submodel; (F) calibration plot comparing the marginal joint-model predicted progression-free survival with the overall Kaplan–Meier estimate.

## Supplementary materials

Supplementary Table 1. Regimen-specific estimates, optimal regimen selection and MCB best-set inclusion for N = 600 and 1200 with a maximum of 25 longitudinal measurements per subject across 1000 replications.

| N | Estimand | DTR | Joint-model framework | | | | IPTW Kaplan–Meier | | | | RE |
|---|---|---|---|---|---|---|---|---|---|---|---|
| | | | Rel% | MCSE | RMSE | Point% | Rel% | MCSE | RMSE | Point% | |
| 600 | $\Psi(16)$ | (A,A,C) | 0.0317 | 0.1951 | 0.1951 | 100.0 | 0.0014 | 0.2569 | 0.2568 | 84.1 | 0.58 |
| | | (A,A,D) | 0.0206 | 0.2072 | 0.2071 | 0.0 | 0.0571 | 0.2576 | 0.2576 | 14.9 | 0.65 |
| | | (B,B,C) | −0.0299 | 0.2189 | 0.2188 | 0.0 | −0.0248 | 0.2832 | 0.2831 | 0.8 | 0.60 |
| | | (B,B,D) | −0.0574 | 0.2300 | 0.2300 | 0.0 | −0.0288 | 0.2811 | 0.2810 | 0.2 | 0.67 |
| | $\Psi(24)$ | (A,A,C) | 0.0193 | 0.4030 | 0.4028 | 100.0 | −0.0402 | 0.5098 | 0.5096 | 98.8 | 0.62 |
| | | (A,A,D) | 0.0306 | 0.4074 | 0.4072 | 0.0 | 0.0830 | 0.4640 | 0.4639 | 1.0 | 0.77 |
| | | (B,B,C) | −0.0301 | 0.4170 | 0.4168 | 0.0 | 0.0224 | 0.5248 | 0.5245 | 0.2 | 0.63 |
| | | (B,B,D) | −0.0695 | 0.3981 | 0.3981 | 0.0 | −0.0181 | 0.4670 | 0.4668 | 0.0 | 0.73 |
| | $S(16)$ | (A,A,C) | 0.0652 | 0.0249 | 0.0249 | 100.0 | 0.0123 | 0.0372 | 0.0372 | 95.0 | 0.45 |
| | | (A,A,D) | 0.0271 | 0.0281 | 0.0281 | 0.0 | 0.3407 | 0.0387 | 0.0388 | 4.6 | 0.53 |
| | | (B,B,C) | −0.0800 | 0.0257 | 0.0257 | 0.0 | −0.0248 | 0.0400 | 0.0400 | 0.4 | 0.41 |
| | | (B,B,D) | −0.2710 | 0.0277 | 0.0277 | 0.0 | 0.1176 | 0.0389 | 0.0389 | 0.0 | 0.51 |
| | $S(24)$ | (A,A,C) | −0.1553 | 0.0369 | 0.0369 | 100.0 | −0.2027 | 0.0396 | 0.0396 | 99.9 | 0.87 |
| | | (A,A,D) | 0.1661 | 0.0268 | 0.0267 | 0.0 | 0.0226 | 0.0342 | 0.0342 | 0.0 | 0.61 |
| | | (B,B,C) | 0.0791 | 0.0349 | 0.0349 | 0.0 | 0.1131 | 0.0376 | 0.0375 | 0.1 | 0.86 |
| | | (B,B,D) | 0.1709 | 0.0208 | 0.0208 | 0.0 | −0.4624 | 0.0280 | 0.0280 | 0.0 | 0.55 |
| 1200 | $\Psi(16)$ | (A,A,C) | 0.0239 | 0.1454 | 0.1453 | 100.0 | −0.0084 | 0.1918 | 0.1917 | 92.7 | 0.57 |
| | | (A,A,D) | 0.0274 | 0.1544 | 0.1544 | 0.0 | 0.0109 | 0.1864 | 0.1863 | 7.3 | 0.69 |
| | | (B,B,C) | 0.0245 | 0.1559 | 0.1558 | 0.0 | 0.0473 | 0.1976 | 0.1976 | 0.0 | 0.62 |
| | | (B,B,D) | 0.0191 | 0.1629 | 0.1628 | 0.0 | 0.0456 | 0.2045 | 0.2045 | 0.0 | 0.63 |
| | $\Psi(24)$ | (A,A,C) | 0.0165 | 0.2973 | 0.2972 | 100.0 | −0.0049 | 0.3767 | 0.3765 | 100.0 | 0.62 |
| | | (A,A,D) | 0.0589 | 0.3032 | 0.3032 | 0.0 | 0.0270 | 0.3414 | 0.3413 | 0.0 | 0.79 |
| | | (B,B,C) | 0.0544 | 0.2983 | 0.2983 | 0.0 | 0.0864 | 0.3675 | 0.3675 | 0.0 | 0.66 |
| | | (B,B,D) | 0.0620 | 0.2845 | 0.2844 | 0.0 | 0.0566 | 0.3293 | 0.3292 | 0.0 | 0.75 |
| | $S(16)$ | (A,A,C) | 0.0489 | 0.0184 | 0.0184 | 100.0 | 0.0264 | 0.0275 | 0.0275 | 99.4 | 0.45 |
| | | (A,A,D) | 0.1046 | 0.0208 | 0.0208 | 0.0 | 0.1232 | 0.0281 | 0.0281 | 0.6 | 0.55 |
| | | (B,B,C) | 0.1130 | 0.0183 | 0.0182 | 0.0 | 0.1475 | 0.0277 | 0.0277 | 0.0 | 0.43 |
| | | (B,B,D) | 0.1173 | 0.0198 | 0.0198 | 0.0 | 0.0412 | 0.0275 | 0.0275 | 0.0 | 0.52 |
| | $S(24)$ | (A,A,C) | −0.0928 | 0.0264 | 0.0264 | 100.0 | −0.0879 | 0.0280 | 0.0280 | 100.0 | 0.89 |
| | | (A,A,D) | 0.3192 | 0.0196 | 0.0196 | 0.0 | 0.0560 | 0.0248 | 0.0248 | 0.0 | 0.63 |
| | | (B,B,C) | 0.2792 | 0.0248 | 0.0248 | 0.0 | 0.1533 | 0.0263 | 0.0263 | 0.0 | 0.89 |
| | | (B,B,D) | 0.7095 | 0.0150 | 0.0150 | 0.0 | 0.2902 | 0.0192 | 0.0192 | 0.0 | 0.61 |

Note: $S(t)$ = survival function at time $t$ = 16, 24 weeks; $\Psi(t^*)$ = restricted mean survival time at time horizon $t^*$ = 16, 24 weeks; DTR = dynamic treatment regimen; Rel% = relative bias in %; MCSE = empirical standard deviation of point estimates across replications; RMSE = root mean squared error; Point % = percentage of simulations selecting each DTR as optimal based on point-estimate; RE = relative efficiency (MCSE-based variance ratio of joint-model framework to IPTW Kaplan-Meier).

Supplementary Table 2. Pairwise regimen contrasts for N = 600 and 1200 with a maximum of 25 longitudinal measurements per subject across 1000 replications.

| N | | 600 | | | | | | | 1200 | | | | | | |
|---|---|---|---|---|---|---|---|---|---|---|---|---|---|---|---|
| Estimand | Contrast | Joint-model framework | | | IPTW Kaplan–Meier | | | RE | Joint-model framework | | | IPTW Kaplan–Meier | | | RE |
| | | Rel% | MCSE | RMSE | Rel% | MCSE | RMSE | | Rel% | MCSE | RMSE | Rel% | MCSE | RMSE | |
| Ψ(16) | (A,A,C)−(A,A,D) | 0.6919 | 0.0513 | 0.0513 | −3.3054 | 0.2192 | 0.2192 | 0.05 | −0.1823 | 0.0355 | 0.0355 | −1.1504 | 0.1563 | 0.1563 | 0.05 |
| | (A,A,C)−(B,B,C) | 0.8980 | 0.2356 | 0.2356 | 0.3695 | 0.3794 | 0.3792 | 0.39 | 0.0158 | 0.1700 | 0.1699 | −0.7911 | 0.2753 | 0.2753 | 0.38 |
| | (A,A,C)−(B,B,D) | 0.9714 | 0.2544 | 0.2546 | 0.3195 | 0.3769 | 0.3768 | 0.46 | 0.0739 | 0.1850 | 0.1849 | −0.5770 | 0.2733 | 0.2732 | 0.46 |
| | (A,A,D)−(B,B,C) | 0.9666 | 0.2515 | 0.2515 | 1.5935 | 0.3753 | 0.3753 | 0.45 | 0.0818 | 0.1815 | 0.1814 | −0.6715 | 0.2719 | 0.2718 | 0.45 |
| | (A,A,D)−(B,B,D) | 1.0375 | 0.2675 | 0.2676 | 1.1774 | 0.3735 | 0.3735 | 0.51 | 0.1345 | 0.1940 | 0.1940 | −0.4413 | 0.2724 | 0.2723 | 0.51 |
| | (B,B,C)−(B,B,D) | 1.2117 | 0.0612 | 0.0612 | 0.1557 | 0.2558 | 0.2557 | 0.06 | 0.2641 | 0.0443 | 0.0443 | 0.1241 | 0.1761 | 0.1760 | 0.06 |
| Ψ(24) | (A,A,C)−(A,A,D) | −0.1255 | 0.2646 | 0.2645 | −1.6147 | 0.4972 | 0.4974 | 0.28 | −0.5253 | 0.1820 | 0.1820 | −0.4122 | 0.3561 | 0.3560 | 0.26 |
| | (A,A,C)−(B,B,C) | 0.3772 | 0.5233 | 0.5231 | −0.4945 | 0.7265 | 0.7262 | 0.52 | −0.2585 | 0.3691 | 0.3689 | −0.6670 | 0.5251 | 0.5250 | 0.49 |
| | (A,A,C)−(B,B,D) | 0.3895 | 0.5327 | 0.5326 | −0.1324 | 0.6861 | 0.6858 | 0.60 | −0.1734 | 0.3878 | 0.3877 | −0.2616 | 0.4893 | 0.4891 | 0.63 |
| | (A,A,D)−(B,B,C) | 1.1272 | 0.5438 | 0.5436 | 1.1767 | 0.6894 | 0.6891 | 0.62 | 0.1395 | 0.3882 | 0.3881 | −1.0472 | 0.4992 | 0.4991 | 0.60 |
| | (A,A,D)−(B,B,D) | 0.6988 | 0.5339 | 0.5338 | 0.7580 | 0.6483 | 0.6482 | 0.68 | 0.0380 | 0.3875 | 0.3873 | −0.1712 | 0.4613 | 0.4611 | 0.71 |
| | (B,B,C)−(B,B,D) | 0.4101 | 0.2697 | 0.2697 | 0.4757 | 0.5262 | 0.5260 | 0.26 | −0.0304 | 0.1964 | 0.1963 | 0.4193 | 0.3607 | 0.3606 | 0.30 |
| $S(16)$ | (A,A,C)−(A,A,D) | 0.3248 | 0.0168 | 0.0168 | −2.2247 | 0.0448 | 0.0448 | 0.14 | −0.3303 | 0.0116 | 0.0116 | −0.6333 | 0.0326 | 0.0326 | 0.13 |
| | (A,A,C)−(B,B,C) | 0.5502 | 0.0339 | 0.0339 | 0.1364 | 0.0547 | 0.0547 | 0.38 | −0.1652 | 0.0239 | 0.0239 | −0.3780 | 0.0390 | 0.0390 | 0.38 |
| | (A,A,C)−(B,B,D) | 0.6259 | 0.0368 | 0.0368 | −0.1632 | 0.0531 | 0.0531 | 0.48 | −0.0651 | 0.0268 | 0.0267 | 0.0017 | 0.0385 | 0.0385 | 0.48 |
| | (A,A,D)−(B,B,C) | 0.8318 | 0.0368 | 0.0368 | 3.0858 | 0.0542 | 0.0542 | 0.46 | 0.0411 | 0.0263 | 0.0263 | −0.0591 | 0.0396 | 0.0396 | 0.44 |
| | (A,A,D)−(B,B,D) | 0.7820 | 0.0381 | 0.0381 | 0.9059 | 0.0538 | 0.0538 | 0.50 | 0.0724 | 0.0276 | 0.0276 | 0.3310 | 0.0391 | 0.0391 | 0.50 |
| | (B,B,C)−(B,B,D) | 0.7467 | 0.0186 | 0.0186 | −0.6413 | 0.0495 | 0.0495 | 0.14 | 0.0946 | 0.0136 | 0.0136 | 0.6078 | 0.0336 | 0.0336 | 0.16 |
| $S(24)$ | (A,A,C)−(A,A,D) | −0.6884 | 0.0351 | 0.0351 | −0.5765 | 0.0443 | 0.0443 | 0.63 | −0.7764 | 0.0241 | 0.0241 | −0.3267 | 0.0314 | 0.0314 | 0.59 |
| | (A,A,C)−(B,B,C) | −0.5800 | 0.0504 | 0.0504 | −0.7750 | 0.0551 | 0.0550 | 0.84 | −0.7672 | 0.0346 | 0.0346 | −0.5250 | 0.0378 | 0.0378 | 0.84 |
| | (A,A,C)−(B,B,D) | −0.3188 | 0.0424 | 0.0423 | −0.0725 | 0.0484 | 0.0484 | 0.77 | −0.4951 | 0.0302 | 0.0303 | −0.2774 | 0.0332 | 0.0332 | 0.83 |
| | (A,A,D)−(B,B,C) | −2.5665 | 0.0436 | 0.0436 | 2.8638 | 0.0514 | 0.0514 | 0.72 | −0.9354 | 0.0307 | 0.0307 | 3.1103 | 0.0353 | 0.0353 | 0.76 |
| | (A,A,D)−(B,B,D) | 0.1606 | 0.0308 | 0.0308 | 0.5811 | 0.0447 | 0.0447 | 0.48 | −0.1303 | 0.0220 | 0.0220 | −0.2135 | 0.0303 | 0.0303 | 0.53 |
| | (B,B,C)−(B,B,D) | −0.0196 | 0.0320 | 0.0320 | 0.7319 | 0.0394 | 0.0394 | 0.66 | −0.1835 | 0.0231 | 0.0231 | 0.0061 | 0.0284 | 0.0284 | 0.66 |

Note: $S(t)$ = survival function at time $t$ = 16, 24 weeks; $\Psi(t^*)$ = restricted mean survival time at time horizon $t^*$ = 16, 24 weeks; Rel% = relative bias in %; MCSE = empirical standard deviation of point estimates across replications; RMSE = root mean squared error; RE = relative efficiency (MCSE-based variance ratio of joint-model framework to IPTW Kaplan-Meier).

Supplementary Table 3. Maximum likelihood estimation of joint-model parameters with a maximum of 4 longitudinal measurements per subject across 1000 replications.

| Parameter | True value | Rel% | MCSE | AESE | RMSE | Cov% | Rel% | MCSE | AESE | RMSE | Cov% | Rel% | MCSE | AESE | RMSE | Cov% |
|---|---|---|---|---|---|---|---|---|---|---|---|---|---|---|---|---|
| | | N=300 | | | | | N=600 | | | | | N=1200 | | | | |
| Longitudinal Submodel | | | | | | | | | | | | | | | | |
| $\beta_0$ | 3.5000 | -0.1521 | 0.0602 | 0.3196 | 0.0604 | 91.4 | -0.1342 | 0.0455 | 0.0868 | 0.0457 | 89.0 | -0.0897 | 0.0316 | 0.0400 | 0.0317 | 90.8 |
| $\beta_{X_{01}}$ | 0.5000 | 0.1545 | 0.0768 | 0.2096 | 0.0768 | 88.1 | -0.3892 | 0.0559 | 0.0619 | 0.0559 | 88.9 | -0.2796 | 0.0393 | 0.0352 | 0.0393 | 89.1 |
| $\beta_{X_{02}}$ | 0.7000 | 0.2191 | 0.0391 | 0.1049 | 0.0392 | 88.8 | -0.2777 | 0.0283 | 0.0277 | 0.0283 | 84.5 | -0.1075 | 0.0190 | 0.0183 | 0.0190 | 89.6 |
| $\beta_t$ | −0.5000 | 0.6749 | 0.1196 | 0.7878 | 0.1196 | 91.7 | 3.4081 | 0.0816 | 0.3127 | 0.0833 | 90.5 | 2.5942 | 0.0610 | 0.1105 | 0.0623 | 90.4 |
| $\beta_A$ | −0.8000 | -0.8364 | 0.1372 | 1.3693 | 0.1373 | 89.6 | -4.3620 | 0.0972 | 0.5293 | 0.1032 | 88.0 | -3.0640 | 0.0728 | 0.1733 | 0.0768 | 86.8 |
| $\beta_B$ | −0.6000 | -1.3020 | 0.1412 | 1.3175 | 0.1413 | 90.8 | -5.5551 | 0.0984 | 0.5389 | 0.1039 | 88.4 | -4.1467 | 0.0737 | 0.1683 | 0.0778 | 87.9 |
| $\beta_C$ | −0.7000 | 0.9285 | 0.1398 | 0.4920 | 0.1399 | 91.7 | 0.6958 | 0.0988 | 0.1689 | 0.0989 | 90.5 | 0.0856 | 0.0711 | 0.0729 | 0.0711 | 91.9 |
| $\sigma_{b_0}$ | 0.5000 | 17.1774 | 0.3656 | 1.5412 | 0.3753 | 68.4 | 10.6299 | 0.0510 | 0.2323 | 0.0736 | 40.1 | 6.4980 | 0.0475 | 0.0844 | 0.0575 | 59.1 |
| $\sigma_{b_1}$ | 0.2000 | -27.8548 | 0.0854 | 2.9522 | 0.1019 | 55.3 | -42.0286 | 0.0697 | 1.2467 | 0.1092 | 34.2 | -27.2307 | 0.0676 | 0.3710 | 0.0868 | 58.9 |
| $\rho$ | −0.3000 | -52.4659 | 0.4687 | 7.0714 | 0.4942 | 54.4 | -119.4210 | 0.3964 | 2.3070 | 0.5342 | 34.8 | -77.9830 | 0.3405 | 0.3941 | 0.4130 | 57.7 |
| $\sigma_\varepsilon$ | 0.5000 | 0.2740 | 0.0241 | 0.5232 | 0.0242 | 87.8 | 2.1692 | 0.0191 | 0.2039 | 0.0219 | 71.2 | 1.4457 | 0.0149 | 0.0627 | 0.0166 | 74.1 |
| Survival Submodel | | | | | | | | | | | | | | | | |
| $\lambda_0$ | 0.1500 | 21.9114 | 0.1085 | 0.2226 | 0.1134 | 93.8 | 18.1255 | 0.0692 | 0.0894 | 0.0743 | 94.1 | 8.9046 | 0.0490 | 0.0551 | 0.0508 | 92.9 |
| $\kappa$ | 2.6000 | 1.2039 | 0.2140 | 0.4071 | 0.2162 | 94.0 | -0.2915 | 0.1395 | 0.1906 | 0.1396 | 95.9 | -0.1142 | 0.1028 | 0.1175 | 0.1028 | 94.8 |
| $\gamma_{X_{01}}$ | 0.4000 | 3.8323 | 0.1965 | 0.3685 | 0.1970 | 95.1 | 6.3948 | 0.1301 | 0.1670 | 0.1325 | 94.7 | 1.5910 | 0.0951 | 0.1073 | 0.0953 | 94.8 |
| $\gamma_{X_{02}}$ | 0.2000 | 8.4136 | 0.1587 | 0.3936 | 0.1595 | 94.2 | 14.9613 | 0.1090 | 0.1574 | 0.1130 | 93.0 | 6.4478 | 0.0784 | 0.0967 | 0.0794 | 94.2 |
| $\gamma_A$ | −0.5000 | -4.7670 | 0.2602 | 0.3253 | 0.2612 | 94.1 | -3.1787 | 0.1814 | 0.1882 | 0.1820 | 94.2 | -0.7273 | 0.1267 | 0.1301 | 0.1267 | 94.0 |
| $\gamma_{AA}$ | −1.5000 | -6.0296 | 0.4914 | 0.6812 | 0.4994 | 96.1 | -4.0111 | 0.3341 | 0.3490 | 0.3393 | 95.4 | -2.4431 | 0.2405 | 0.2538 | 0.2432 | 95.0 |
| $\gamma_{BB}$ | −1.4000 | -6.2090 | 0.4787 | 0.6163 | 0.4863 | 96.5 | -3.3653 | 0.3264 | 0.3338 | 0.3296 | 94.7 | -1.9799 | 0.2233 | 0.2434 | 0.2249 | 95.3 |
| $\gamma_{AC}$ | −1.0000 | -4.9694 | 0.4133 | 0.6170 | 0.4161 | 96.7 | -3.6385 | 0.2848 | 0.3286 | 0.2870 | 95.1 | -0.9647 | 0.2045 | 0.2155 | 0.2046 | 95.0 |
| $\gamma_{BC}$ | −0.9000 | -4.1937 | 0.3766 | 0.5775 | 0.3783 | 95.7 | -4.8550 | 0.2661 | 0.3029 | 0.2696 | 94.6 | -1.6862 | 0.1882 | 0.2049 | 0.1888 | 94.1 |
| $\alpha_m$ | 0.2000 | -8.8145 | 0.1937 | 0.5304 | 0.1944 | 95.2 | -17.9731 | 0.1323 | 0.2050 | 0.1370 | 93.8 | -7.6594 | 0.0993 | 0.1233 | 0.1005 | 93.8 |

Note: Rel% = relative bias in %; MCSE = empirical standard deviation of point estimates across replications; AESE = the average of estimated standard errors from the inverse observed information matrix; RMSE = root mean squared error; Cov% = coverage probability of the nominal 95% Wald confidence interval based on AESE.

Supplementary Table 4: Regimen-specific estimates, optimal regimen selection and MCB best-set inclusion for N = 300 with a maximum of 4 longitudinal measurements per subject across 1000 replications.

| Estimand | DTR | True value | Joint-model framework | | | | | | | IPTW Kaplan–Meier | | | | | | | RE |
|---|---|---|---|---|---|---|---|---|---|---|---|---|---|---|---|---|---|
| | | | Rel% | MCSE | AESE | RMSE | Cov% | Point% | MCB% | Rel% | MCSE | AESE | RMSE | Cov% | Point% | MCB% | |
| $\Psi(16)$ | (A,A,C) | 13.3525 | 0.1352 | 0.2889 | 0.3688 | 0.2893 | 92.6 | 99.2 | 100.0 | 0.1243 | 0.3784 | 0.3813 | 0.3786 | 94.5 | 73.5 | 99.9 | 0.58 |
| | (A,A,D) | 13.1311 | 0.1055 | 0.3045 | 0.3850 | 0.3047 | 93.0 | 0.0 | 10.0 | 0.1467 | 0.3702 | 0.3790 | 0.3705 | 94.4 | 22.1 | 93.9 | 0.68 |
| | (B,B,C) | 12.4664 | -0.0399 | 0.3196 | 0.3814 | 0.3195 | 93.1 | 0.8 | 24.8 | -0.0757 | 0.4080 | 0.4137 | 0.4079 | 95.0 | 3.7 | 62.7 | 0.61 |
| | (B,B,D) | 12.1957 | -0.0767 | 0.3344 | 0.3950 | 0.3344 | 92.5 | 0.0 | 2.6 | -0.1109 | 0.3986 | 0.4031 | 0.3986 | 94.6 | 0.7 | 39.9 | 0.70 |
| $\Psi(24)$ | (A,A,C) | 17.5462 | 0.1935 | 0.5900 | 0.6922 | 0.5906 | 93.4 | 99.5 | 100.0 | 0.1564 | 0.7419 | 0.7386 | 0.7420 | 94.3 | 94.4 | 100.0 | 0.63 |
| | (A,A,D) | 16.2729 | 0.1729 | 0.5873 | 0.6970 | 0.5877 | 93.9 | 0.0 | 8.3 | 0.1536 | 0.6852 | 0.6859 | 0.6853 | 94.7 | 3.6 | 63.4 | 0.73 |
| | (B,B,C) | 15.4195 | -0.0580 | 0.6065 | 0.6802 | 0.6063 | 93.0 | 0.5 | 18.6 | -0.0582 | 0.7553 | 0.7514 | 0.7550 | 94.9 | 2.0 | 47.3 | 0.64 |
| | (B,B,D) | 14.1533 | -0.0567 | 0.5796 | 0.6539 | 0.5793 | 92.6 | 0.0 | 0.8 | -0.2067 | 0.6529 | 0.6505 | 0.6533 | 94.6 | 0.0 | 6.5 | 0.79 |
| $S(16)$ | (A,A,C) | 0.5994 | 0.4063 | 0.0363 | 0.0404 | 0.0364 | 93.9 | 99.5 | 100.0 | 0.2215 | 0.0537 | 0.0531 | 0.0537 | 94.2 | 85.8 | 99.9 | 0.46 |
| | (A,A,D) | 0.5227 | 0.2849 | 0.0401 | 0.0444 | 0.0401 | 93.8 | 0.0 | 9.3 | 0.3191 | 0.0553 | 0.0553 | 0.0553 | 94.2 | 11.9 | 79.2 | 0.52 |
| | (B,B,C) | 0.4613 | -0.1683 | 0.0372 | 0.0399 | 0.0372 | 93.2 | 0.5 | 18.5 | 0.1708 | 0.0555 | 0.0558 | 0.0555 | 94.6 | 2.3 | 56.6 | 0.45 |
| | (B,B,D) | 0.3747 | -0.3707 | 0.0397 | 0.0421 | 0.0397 | 93.7 | 0.0 | 1.1 | -0.8609 | 0.0558 | 0.0544 | 0.0559 | 93.4 | 0.0 | 13.9 | 0.51 |
| $S(24)$ | (A,A,C) | 0.4659 | 0.2675 | 0.0529 | 0.0558 | 0.0529 | 93.2 | 98.9 | 100.0 | 0.1715 | 0.0565 | 0.0566 | 0.0565 | 94.3 | 98.4 | 100.0 | 0.88 |
| | (A,A,D) | 0.2907 | 0.7463 | 0.0382 | 0.0425 | 0.0382 | 94.7 | 0.0 | 7.6 | 0.4493 | 0.0498 | 0.0508 | 0.0498 | 95.1 | 0.3 | 24.7 | 0.59 |
| | (B,B,C) | 0.3003 | 0.0095 | 0.0504 | 0.0518 | 0.0504 | 94.3 | 1.1 | 35.1 | -0.1086 | 0.0542 | 0.0536 | 0.0542 | 93.8 | 1.3 | 45.5 | 0.86 |
| | (B,B,D) | 0.1556 | 1.0943 | 0.0309 | 0.0342 | 0.0309 | 95.1 | 0.0 | 0.6 | -0.8804 | 0.0397 | 0.0395 | 0.0397 | 94.3 | 0.0 | 0.8 | 0.61 |

Note: $\boldsymbol{S(t)}$ = survival function at time $t$ = 16, 24 weeks; $\Psi(t^*)$ = restricted mean survival time at time horizon $t^*$ = 16, 24 weeks; DTR = dynamic treatment regimen; Rel% = relative bias in %; MCSE = empirical standard deviation of point estimates across replications; AESE = the average of estimated standard errors from the multivariate normal propagation for joint-model framework and the bootstrap replicates for IPTW estimator; RMSE = root mean squared error; Cov% = coverage probability of the nominal 95% Wald confidence interval based on AESE; Point % = percentage of simulations selecting each DTR as optimal based on point-estimate; MCB% = percentage of simulations including each DTR in the MCB best set; RE = relative efficiency (MCSE-based variance ratio of joint-model framework to IPTW Kaplan-Meier).

Supplementary Table 5. Pairwise regimen contrasts for N = 300 with a maximum of 4 longitudinal measurements per subject across 1000 replications.

| Estimand | Contrast | True value | Joint-model framework | | | | | IPTW Kaplan–Meier | | | | | RE |
|---|---|---|---|---|---|---|---|---|---|---|---|---|---|
| | | | Rel% | MCSE | AESE | RMSE | Cov% | Rel% | MCSE | AESE | RMSE | Cov% | |
| Ψ(16) | (A,A,C)−(A,A,D) | 0.2214 | 1.9001 | 0.0718 | 0.0770 | 0.0718 | 94.7 | -1.2078 | 0.3157 | 0.3452 | 0.3156 | 96.2 | 0.05 |
| | (A,A,C)−(B,B,C) | 0.8861 | 2.5993 | 0.3475 | 0.3541 | 0.3481 | 93.1 | 2.9379 | 0.5497 | 0.5626 | 0.5501 | 94.8 | 0.40 |
| | (A,A,C)−(B,B,D) | 1.1569 | 2.3698 | 0.3772 | 0.3845 | 0.3780 | 92.0 | 2.6030 | 0.5529 | 0.5554 | 0.5535 | 94.7 | 0.47 |
| | (A,A,D)−(B,B,C) | 0.6647 | 2.8321 | 0.3723 | 0.3784 | 0.3726 | 92.9 | 4.3186 | 0.5484 | 0.5609 | 0.5489 | 94.9 | 0.46 |
| | (A,A,D)−(B,B,D) | 0.9355 | 2.4810 | 0.3979 | 0.4012 | 0.3983 | 92.4 | 3.5048 | 0.5454 | 0.5533 | 0.5461 | 94.9 | 0.53 |
| | (B,B,C)−(B,B,D) | 0.2707 | 1.6188 | 0.0894 | 0.0937 | 0.0895 | 92.7 | 1.5068 | 0.3483 | 0.4027 | 0.3482 | 97.3 | 0.07 |
| Ψ(24) | (A,A,C)−(A,A,D) | 1.2732 | 0.4571 | 0.3687 | 0.3781 | 0.3686 | 93.8 | 0.1922 | 0.7336 | 0.7316 | 0.7332 | 95.2 | 0.25 |
| | (A,A,C)−(B,B,C) | 2.1267 | 2.0171 | 0.7587 | 0.7610 | 0.7596 | 93.7 | 1.7122 | 1.0376 | 1.0527 | 1.0377 | 95.6 | 0.53 |
| | (A,A,C)−(B,B,D) | 3.3929 | 1.2375 | 0.7916 | 0.7958 | 0.7923 | 92.0 | 1.6709 | 0.9985 | 0.9849 | 0.9997 | 94.6 | 0.63 |
| | (A,A,D)−(B,B,C) | 0.8535 | 4.3445 | 0.8045 | 0.7968 | 0.8050 | 93.0 | 3.9799 | 1.0302 | 1.0171 | 1.0302 | 94.2 | 0.61 |
| | (A,A,D)−(B,B,D) | 2.1196 | 1.7063 | 0.8053 | 0.7956 | 0.8057 | 92.2 | 2.5591 | 0.9522 | 0.9458 | 0.9533 | 94.7 | 0.72 |
| | (B,B,C)−(B,B,D) | 1.2661 | -0.0721 | 0.3913 | 0.3955 | 0.3911 | 92.9 | 1.6014 | 0.7190 | 0.7553 | 0.7189 | 95.3 | 0.30 |
| $S(16)$ | (A,A,C)−(A,A,D) | 0.0767 | 1.2337 | 0.0234 | 0.0248 | 0.0234 | 94.0 | -0.4432 | 0.0667 | 0.0642 | 0.0666 | 93.5 | 0.12 |
| | (A,A,C)−(B,B,C) | 0.1382 | 2.3248 | 0.0494 | 0.0493 | 0.0495 | 93.6 | 0.3909 | 0.0749 | 0.0770 | 0.0748 | 96.2 | 0.44 |
| | (A,A,C)−(B,B,D) | 0.2247 | 1.7020 | 0.0546 | 0.0547 | 0.0547 | 92.5 | 2.0266 | 0.0784 | 0.0761 | 0.0785 | 92.7 | 0.48 |
| | (A,A,D)−(B,B,C) | 0.0614 | 3.6878 | 0.0544 | 0.0541 | 0.0544 | 93.1 | 1.4327 | 0.0785 | 0.0786 | 0.0784 | 94.4 | 0.48 |
| | (A,A,D)−(B,B,D) | 0.1480 | 1.9448 | 0.0572 | 0.0561 | 0.0573 | 92.3 | 3.3073 | 0.0791 | 0.0777 | 0.0793 | 95.0 | 0.52 |
| | (B,B,C)−(B,B,D) | 0.0866 | 0.7077 | 0.0272 | 0.0283 | 0.0272 | 92.4 | 4.6377 | 0.0682 | 0.0679 | 0.0683 | 94.6 | 0.16 |
| $S(24)$ | (A,A,C)−(A,A,D) | 0.1752 | -0.5269 | 0.0490 | 0.0490 | 0.0489 | 93.1 | -0.2893 | 0.0639 | 0.0636 | 0.0639 | 94.6 | 0.59 |
| | (A,A,C)−(B,B,C) | 0.1657 | 0.7349 | 0.0702 | 0.0703 | 0.0701 | 95.1 | 0.6793 | 0.0760 | 0.0779 | 0.0760 | 95.6 | 0.85 |
| | (A,A,C)−(B,B,D) | 0.3103 | -0.1470 | 0.0613 | 0.0622 | 0.0613 | 92.2 | 0.6989 | 0.0697 | 0.0692 | 0.0697 | 94.7 | 0.77 |
| | (A,A,D)−(B,B,C) | −0.0096 | -22.3927 | 0.0636 | 0.0621 | 0.0636 | 93.3 | -17.0752 | 0.0756 | 0.0739 | 0.0756 | 93.9 | 0.71 |
| | (A,A,D)−(B,B,D) | 0.1351 | 0.3456 | 0.0474 | 0.0471 | 0.0474 | 93.4 | 1.9805 | 0.0654 | 0.0646 | 0.0654 | 94.3 | 0.53 |
| | (B,B,C)−(B,B,D) | 0.1447 | -1.1569 | 0.0458 | 0.0454 | 0.0458 | 92.7 | 0.7213 | 0.0566 | 0.0563 | 0.0565 | 93.7 | 0.65 |

Note: $S(t)$ = survival function at time $t$ = 16, 24 weeks; $\Psi(t^*)$ = restricted mean survival time at time horizon $t^*$ = 16, 24 weeks; Rel% = relative bias in %; MCSE = empirical standard deviation of point estimates across replications; AESE = the average of estimated standard errors from the multivariate normal propagation for joint-model framework and the bootstrap replicates for IPTW estimator; Cov% = coverage probability of the nominal 95% Wald confidence interval based on AESE; RE = relative efficiency (MCSE-based variance ratio of joint-model framework to IPTW Kaplan-Meier).

Supplementary Table 6. Regimen-specific estimates and optimal regimen selection for N = 600 and 1200 with a maximum of 4 longitudinal measurements per subject across 1000 replications.

| N | Estimand | DTR | Joint-model framework | | | | IPTW Kaplan–Meier | | | | RE |
|---|---|---|---|---|---|---|---|---|---|---|---|
| | | | Rel% | MCSE | RMSE | Point% | Rel% | MCSE | RMSE | Point% | |
| 600 | $\Psi(16)$ | (A,A,C) | 0.0784 | 0.1966 | 0.1968 | 99.9 | 0.0065 | 0.2598 | 0.2597 | 83.6 | 0.57 |
| | | (A,A,D) | 0.0981 | 0.2106 | 0.2109 | 0.0 | 0.0299 | 0.2576 | 0.2575 | 15.5 | 0.67 |
| | | (B,B,C) | 0.0234 | 0.2182 | 0.2182 | 0.0 | −0.0305 | 0.2874 | 0.2873 | 0.6 | 0.58 |
| | | (B,B,D) | 0.0228 | 0.2271 | 0.2270 | 0.0 | −0.0210 | 0.2779 | 0.2778 | 0.3 | 0.67 |
| | $\Psi(24)$ | (A,A,C) | 0.1365 | 0.4004 | 0.4009 | 100.0 | −0.0147 | 0.5075 | 0.5072 | 98.4 | 0.62 |
| | | (A,A,D) | 0.2688 | 0.4198 | 0.4219 | 0.0 | 0.0888 | 0.4881 | 0.4881 | 1.5 | 0.74 |
| | | (B,B,C) | 0.0964 | 0.4245 | 0.4246 | 0.0 | 0.0033 | 0.5326 | 0.5323 | 0.1 | 0.64 |
| | | (B,B,D) | 0.1397 | 0.3982 | 0.3985 | 0.0 | −0.0348 | 0.4557 | 0.4555 | 0.0 | 0.76 |
| | $S(16)$ | (A,A,C) | 0.3121 | 0.0248 | 0.0248 | 100.0 | −0.0261 | 0.0373 | 0.0373 | 94.5 | 0.44 |
| | | (A,A,D) | 0.5544 | 0.0288 | 0.0289 | 0.0 | 0.2458 | 0.0394 | 0.0394 | 5.3 | 0.54 |
| | | (B,B,C) | 0.2386 | 0.0261 | 0.0261 | 0.0 | 0.0875 | 0.0398 | 0.0398 | 0.2 | 0.43 |
| | | (B,B,D) | 0.3516 | 0.0277 | 0.0277 | 0.0 | 0.1446 | 0.0382 | 0.0382 | 0.0 | 0.52 |
| | $S(24)$ | (A,A,C) | 0.2605 | 0.0363 | 0.0363 | 100.0 | −0.0886 | 0.0387 | 0.0387 | 99.8 | 0.88 |
| | | (A,A,D) | 1.5257 | 0.0282 | 0.0285 | 0.0 | 0.3499 | 0.0359 | 0.0359 | 0.0 | 0.61 |
| | | (B,B,C) | 0.6107 | 0.0363 | 0.0364 | 0.0 | 0.2363 | 0.0390 | 0.0390 | 0.2 | 0.87 |
| | | (B,B,D) | 1.6810 | 0.0220 | 0.0221 | 0.0 | 0.0756 | 0.0282 | 0.0282 | 0.0 | 0.61 |
| 1200 | $\Psi(16)$ | (A,A,C) | 0.0444 | 0.1417 | 0.1418 | 100.0 | −0.0108 | 0.1880 | 0.1880 | 91.7 | 0.57 |
| | | (A,A,D) | 0.0641 | 0.1507 | 0.1508 | 0.0 | 0.0020 | 0.1894 | 0.1893 | 8.3 | 0.63 |
| | | (B,B,C) | 0.0675 | 0.1577 | 0.1579 | 0.0 | 0.0461 | 0.2052 | 0.2052 | 0.0 | 0.59 |
| | | (B,B,D) | 0.0807 | 0.1648 | 0.1650 | 0.0 | 0.0610 | 0.1989 | 0.1989 | 0.0 | 0.69 |
| | $\Psi(24)$ | (A,A,C) | 0.0672 | 0.2964 | 0.2965 | 100.0 | 0.0186 | 0.3698 | 0.3696 | 100.0 | 0.64 |
| | | (A,A,D) | 0.1748 | 0.3031 | 0.3043 | 0.0 | 0.0638 | 0.3509 | 0.3509 | 0.0 | 0.75 |
| | | (B,B,C) | 0.1333 | 0.3031 | 0.3036 | 0.0 | 0.0600 | 0.3849 | 0.3848 | 0.0 | 0.62 |
| | | (B,B,D) | 0.1986 | 0.2883 | 0.2896 | 0.0 | 0.0766 | 0.3248 | 0.3248 | 0.0 | 0.79 |
| | $S(16)$ | (A,A,C) | 0.1515 | 0.0183 | 0.0183 | 100.0 | 0.0814 | 0.0263 | 0.0263 | 99.0 | 0.48 |
| | | (A,A,D) | 0.3517 | 0.0207 | 0.0207 | 0.0 | 0.0542 | 0.0285 | 0.0285 | 1.0 | 0.53 |
| | | (B,B,C) | 0.3002 | 0.0186 | 0.0187 | 0.0 | 0.0445 | 0.0291 | 0.0291 | 0.0 | 0.41 |
| | | (B,B,D) | 0.5065 | 0.0199 | 0.0200 | 0.0 | 0.1564 | 0.0279 | 0.0279 | 0.0 | 0.51 |
| | $S(24)$ | (A,A,C) | 0.0908 | 0.0271 | 0.0271 | 100.0 | −0.0237 | 0.0290 | 0.0290 | 100.0 | 0.88 |
| | | (A,A,D) | 1.0192 | 0.0207 | 0.0209 | 0.0 | 0.1997 | 0.0258 | 0.0257 | 0.0 | 0.64 |
| | | (B,B,C) | 0.5362 | 0.0254 | 0.0255 | 0.0 | 0.2987 | 0.0276 | 0.0276 | 0.0 | 0.85 |
| | | (B,B,D) | 1.5815 | 0.0157 | 0.0159 | 0.0 | 0.1904 | 0.0200 | 0.0200 | 0.0 | 0.61 |

Note: $S(t)$ = survival function at time $t$ = 16, 24 weeks; $\Psi(t^*)$ = restricted mean survival time at time horizon $t^*$ = 16, 24 weeks; DTR = dynamic treatment regimen; Rel% = relative bias in %; MCSE = empirical standard deviation of point estimates across replications; RMSE = root mean squared error; Point % = percentage of simulations selecting each DTR as optimal based on point-estimate; RE = relative efficiency (MCSE-based variance ratio of joint-model framework to IPTW Kaplan-Meier).

Supplementary Table 7. Pairwise regimen contrasts for N = 600 and 1200 with a maximum of 4 longitudinal measurements per subject across 1000 replications.

| N | | 600 | | | | | | | 1200 | | | | | | |
|---|---|---|---|---|---|---|---|---|---|---|---|---|---|---|---|
| Estimand | Contrast | Joint-model framework | | | IPTW Kaplan–Meier | | | RE | Joint-model framework | | | IPTW Kaplan–Meier | | | RE |
| | | Rel% | MCSE | RMSE | Rel% | MCSE | RMSE | | Rel% | MCSE | RMSE | Rel% | MCSE | RMSE | |
| Ψ(16) | (A,A,C)−(A,A,D) | −1.0871 | 0.0511 | 0.0512 | −1.3855 | 0.2255 | 0.2254 | 0.05 | −1.1220 | 0.0367 | 0.0367 | −0.7706 | 0.1605 | 0.1605 | 0.05 |
| | (A,A,C)−(B,B,C) | 0.8523 | 0.2389 | 0.2389 | 0.5261 | 0.3766 | 0.3765 | 0.40 | −0.2793 | 0.1690 | 0.1689 | −0.8106 | 0.2751 | 0.2751 | 0.38 |
| | (A,A,C)−(B,B,D) | 0.6654 | 0.2581 | 0.2581 | 0.2958 | 0.3761 | 0.3759 | 0.47 | −0.3380 | 0.1831 | 0.1831 | −0.7676 | 0.2686 | 0.2686 | 0.46 |
| | (A,A,D)−(B,B,C) | 1.4983 | 0.2583 | 0.2584 | 1.1628 | 0.3852 | 0.3851 | 0.45 | 0.0013 | 0.1822 | 0.1821 | −0.8240 | 0.2794 | 0.2793 | 0.43 |
| | (A,A,D)−(B,B,D) | 1.0801 | 0.2745 | 0.2746 | 0.6937 | 0.3794 | 0.3793 | 0.52 | −0.1525 | 0.1942 | 0.1941 | −0.7668 | 0.2734 | 0.2733 | 0.50 |
| | (B,B,C)−(B,B,D) | 0.0534 | 0.0630 | 0.0630 | −0.4580 | 0.2548 | 0.2547 | 0.06 | −0.5301 | 0.0441 | 0.0441 | −0.6266 | 0.1749 | 0.1748 | 0.06 |
| Ψ(24) | (A,A,C)−(A,A,D) | −1.5549 | 0.2634 | 0.2640 | −1.3381 | 0.5295 | 0.5295 | 0.25 | −1.3088 | 0.1900 | 0.1907 | −0.5586 | 0.3697 | 0.3696 | 0.26 |
| | (A,A,C)−(B,B,C) | 0.4270 | 0.5200 | 0.5198 | −0.1458 | 0.7019 | 0.7016 | 0.55 | −0.4121 | 0.3702 | 0.3701 | −0.2809 | 0.5220 | 0.5218 | 0.50 |
| | (A,A,C)−(B,B,D) | 0.1230 | 0.5353 | 0.5350 | 0.0689 | 0.6718 | 0.6715 | 0.63 | −0.4811 | 0.3851 | 0.3853 | −0.2230 | 0.4835 | 0.4833 | 0.63 |
| | (A,A,D)−(B,B,C) | 3.3838 | 0.5628 | 0.5633 | 1.6329 | 0.7258 | 0.7256 | 0.60 | 0.9256 | 0.3965 | 0.3964 | 0.1333 | 0.5227 | 0.5224 | 0.58 |
| | (A,A,D)−(B,B,D) | 1.1310 | 0.5564 | 0.5566 | 0.9142 | 0.6681 | 0.6680 | 0.69 | 0.0161 | 0.3941 | 0.3939 | −0.0214 | 0.4787 | 0.4784 | 0.68 |
| | (B,B,C)−(B,B,D) | −0.3876 | 0.2799 | 0.2798 | 0.4297 | 0.5304 | 0.5302 | 0.28 | −0.5970 | 0.1973 | 0.1974 | −0.1258 | 0.3753 | 0.3751 | 0.28 |
| $S(16)$ | (A,A,C)−(A,A,D) | −1.3382 | 0.0167 | 0.0168 | −1.8779 | 0.0457 | 0.0457 | 0.13 | −1.2125 | 0.0120 | 0.0120 | 0.2663 | 0.0331 | 0.0331 | 0.13 |
| | (A,A,C)−(B,B,C) | 0.5576 | 0.0338 | 0.0338 | −0.4051 | 0.0526 | 0.0526 | 0.41 | −0.3452 | 0.0240 | 0.0240 | 0.2047 | 0.0384 | 0.0384 | 0.39 |
| | (A,A,C)−(B,B,D) | 0.2463 | 0.0370 | 0.0370 | −0.3107 | 0.0525 | 0.0524 | 0.50 | −0.4405 | 0.0265 | 0.0265 | −0.0436 | 0.0375 | 0.0374 | 0.50 |
| | (A,A,D)−(B,B,C) | 2.9257 | 0.0380 | 0.0381 | 1.4346 | 0.0563 | 0.0563 | 0.46 | 0.7381 | 0.0267 | 0.0267 | 0.1276 | 0.0411 | 0.0411 | 0.42 |
| | (A,A,D)−(B,B,D) | 1.0679 | 0.0395 | 0.0395 | 0.5020 | 0.0545 | 0.0545 | 0.53 | −0.0403 | 0.0280 | 0.0280 | −0.2043 | 0.0408 | 0.0408 | 0.47 |
| | (B,B,C)−(B,B,D) | −0.2507 | 0.0193 | 0.0193 | −0.1599 | 0.0482 | 0.0482 | 0.16 | −0.5927 | 0.0135 | 0.0135 | −0.4400 | 0.0345 | 0.0345 | 0.15 |
| $S(24)$ | (A,A,C)−(A,A,D) | −1.8384 | 0.0348 | 0.0350 | −0.8161 | 0.0466 | 0.0466 | 0.56 | −1.4492 | 0.0253 | 0.0254 | −0.3942 | 0.0326 | 0.0326 | 0.60 |
| | (A,A,C)−(B,B,C) | −0.3741 | 0.0492 | 0.0492 | −0.6776 | 0.0531 | 0.0530 | 0.86 | −0.7165 | 0.0355 | 0.0355 | −0.6080 | 0.0387 | 0.0387 | 0.84 |
| | (A,A,C)−(B,B,D) | −0.4516 | 0.0420 | 0.0420 | −0.1710 | 0.0480 | 0.0480 | 0.76 | −0.6565 | 0.0311 | 0.0312 | −0.1310 | 0.0352 | 0.0352 | 0.78 |
| | (A,A,D)−(B,B,C) | −27.2136 | 0.0452 | 0.0453 | −3.2174 | 0.0532 | 0.0532 | 0.72 | −14.1477 | 0.0320 | 0.0320 | 3.3098 | 0.0370 | 0.0370 | 0.75 |
| | (A,A,D)−(B,B,D) | 1.3470 | 0.0333 | 0.0333 | 0.6657 | 0.0460 | 0.0460 | 0.52 | 0.3716 | 0.0235 | 0.0235 | 0.2104 | 0.0324 | 0.0324 | 0.53 |
| | (B,B,C)−(B,B,D) | −0.5403 | 0.0333 | 0.0333 | 0.4091 | 0.0412 | 0.0412 | 0.65 | −0.5879 | 0.0236 | 0.0236 | 0.4152 | 0.0297 | 0.0297 | 0.63 |

Note: $S(t)$ = survival function at time $t$ = 16, 24 weeks; $\Psi(t^*)$ = restricted mean survival time at time horizon $t^*$ = 16, 24 weeks; Rel% = relative bias in %; MCSE = empirical standard deviation of point estimates across replications; RMSE = root mean squared error; RE = relative efficiency (MCSE-based variance ratio of joint-model framework to IPTW Kaplan-Meier).

Supplementary Table 8. Distribution of patients by first-stage treatment and embedded dynamic treatment regimens in the androgen-independent prostate cancer trial.

| $V_1$ | $N_{V_1}$ | $N_{off}$ | $N_{e<8}$ | DTR | $N_R$ | $N_{e,R}$ | $N_{NR}$ | $N_{e,NR}$ |
|---|---|---|---|---|---|---|---|---|
| TEC | 38 | 3 | 1(2.6%) | (TEC,TEC,TEE) | 25 (65.8%) | 2 | 3 | 3 |
| | | | | (TEC,TEC,KA/VE) | | | 4 | 0 |
| | | | | (TEC,TEC,CVD) | | | 2 | 1 |
| TEE | 39 | 5 | 4(10.3%) | (TEE,TEE,TEC) | 20(51.3%) | 3 | 4 | 3 |
| | | | | (TEE,TEE,KA/VE) | | | 3 | 2 |
| | | | | (TEE,TEE,CVD) | | | 3 | 1 |
| KA/VE | 36 | 2 | 4(11.1%) | (KA/VE,KA/VE,TEC) | 24 (66.7%) | 5 | 1 | 0 |
| | | | | (KA/VE,KA/VE,TEE) | | | 3 | 2 |
| | | | | (KA/VE,KA/VE,CVD) | | | 2 | 0 |
| CVD | 37 | 1 | 11(29.7%) | (CVD,CVD,TEC) | 10 (27.0%) | 2 | 5 | 0 |
| | | | | (CVD,CVD,TEE) | | | 2 | 0 |
| | | | | (CVD,CVD,KA/VE) | | | 8 | 1 |

Note: $V_1$= first-stage treatment; $N_{V_1}$= number of patients randomized to each first-stage treatment; $N_{off}$ = number of patients who went off protocol at week 8; $N_{e<8}$ = number of patients who progressed before week 8; DTR = dynamic treatment regimen; $N_R$ = number of responders for each $V_1$; $N_{e,R}$ = number of progressions among responders; $N_{NR}$ = number of non-responders randomized to the second-stage treatment of each DTR; $N_{e,NR}$ = number of progression among non-responders. Percentages for $N_{e<8}$ and $N_R$ are computed relative to $N_{V_1}$

Supplementary Table 9. Regimen-specific progression-free survival probability estimates at week 16 and MCB best-set inclusion for the androgen-independent prostate cancer trial.

| Joint-model framework | | | | IPTW Kaplan–Meier | | | |
|---|---|---|---|---|---|---|---|
| DTR | $S(16)$ (95% CI) | $\mathbb{M}$ | In best set | DTR | $S(16)$ (95% CI) | $\mathbb{M}$ | In best set |
| (TEC,TEC,KA/VE) | 0.8425 (0.7178, 0.9415) | 2.6834 | Yes | (TEC,TEC,KA/VE) | 0.9254 (0.8295, 1.0000) | 3.8610 | Yes |
| (TEC,TEC,TEE) | 0.8295 (0.6951, 0.9359) | 2.0915 | Yes | (TEC,TEC,CVD) | 0.8187 (0.6217, 1.0000) | 1.1364 | Yes |
| (KA/VE,KA/VE,TEC) | 0.8053 (0.6653, 0.9028) | 1.7875 | Yes | (KA/VE,KA/VE,CVD) | 0.7500 (0.5975, 0.8925) | 0.4529 | Yes |
| (KA/VE,KA/VE,TEE) | 0.7805 (0.6300, 0.8848) | 1.4905 | Yes | (KA/VE,KA/VE,TEC) | 0.7290 (0.5586, 0.8806) | 0.2799 | Yes |
| (TEE,TEE,TEC) | 0.7781 (0.6304, 0.8869) | 1.5077 | Yes | (TEE,TEE,CVD) | 0.7226 (0.4985, 0.9048) | 0.6061 | Yes |
| (TEE,TEE,KA/VE) | 0.7455 (0.5983, 0.8644) | 1.0561 | Yes | (CVD,CVD,TEC) | 0.7047 (0.5018, 0.8444) | 0.2717 | Yes |
| (TEC,TEC,CVD) | 0.7419 (0.5968, 0.9137) | 0.4578 | Yes | (CVD,CVD,KA/VE) | 0.6759 (0.4692, 0.8328) | 0.0488 | Yes |
| (KA/VE,KA/VE,CVD) | 0.7353 (0.5900, 0.8563) | 0.7445 | Yes | (TEC,TEC,TEE) | 0.6611 (0.4540, 0.9032) | -0.0527 | No |
| (CVD,CVD,TEC) | 0.6921 (0.5158, 0.7984) | 0.6747 | Yes | (CVD,CVD,TEE) | 0.6402 (0.4035, 0.8139) | -0.0763 | No |
| (CVD,CVD,KA/VE) | 0.6732 (0.4920, 0.7851) | 0.4141 | Yes | (TEE,TEE,KA/VE) | 0.6284 (0.4014, 0.8495) | -0.1316 | No |
| (CVD,CVD,TEE) | 0.6664 (0.4804, 0.7835) | 0.4272 | Yes | (KA/VE,KA/VE,TEE) | 0.6032 (0.4156, 0.8007) | -0.5114 | No |
| (TEE,TEE,CVD) | 0.6556 (0.5142, 0.8178) | 0.1664 | Yes | (TEE,TEE,TEC) | 0.5718 (0.3602, 0.8189) | -0.4799 | No |

Note: DTR = dynamic treatment regimen; $S_{AIPC}(16)$ = progression-free survival probability at week 16; MCB = multiple comparisons with the best; $\mathbb{M}$ = multiplicity-adjusted MCB inclusion margin (positive = included, negative = excluded). MCB is implemented at significance level $\varsigma$ = 0.05.

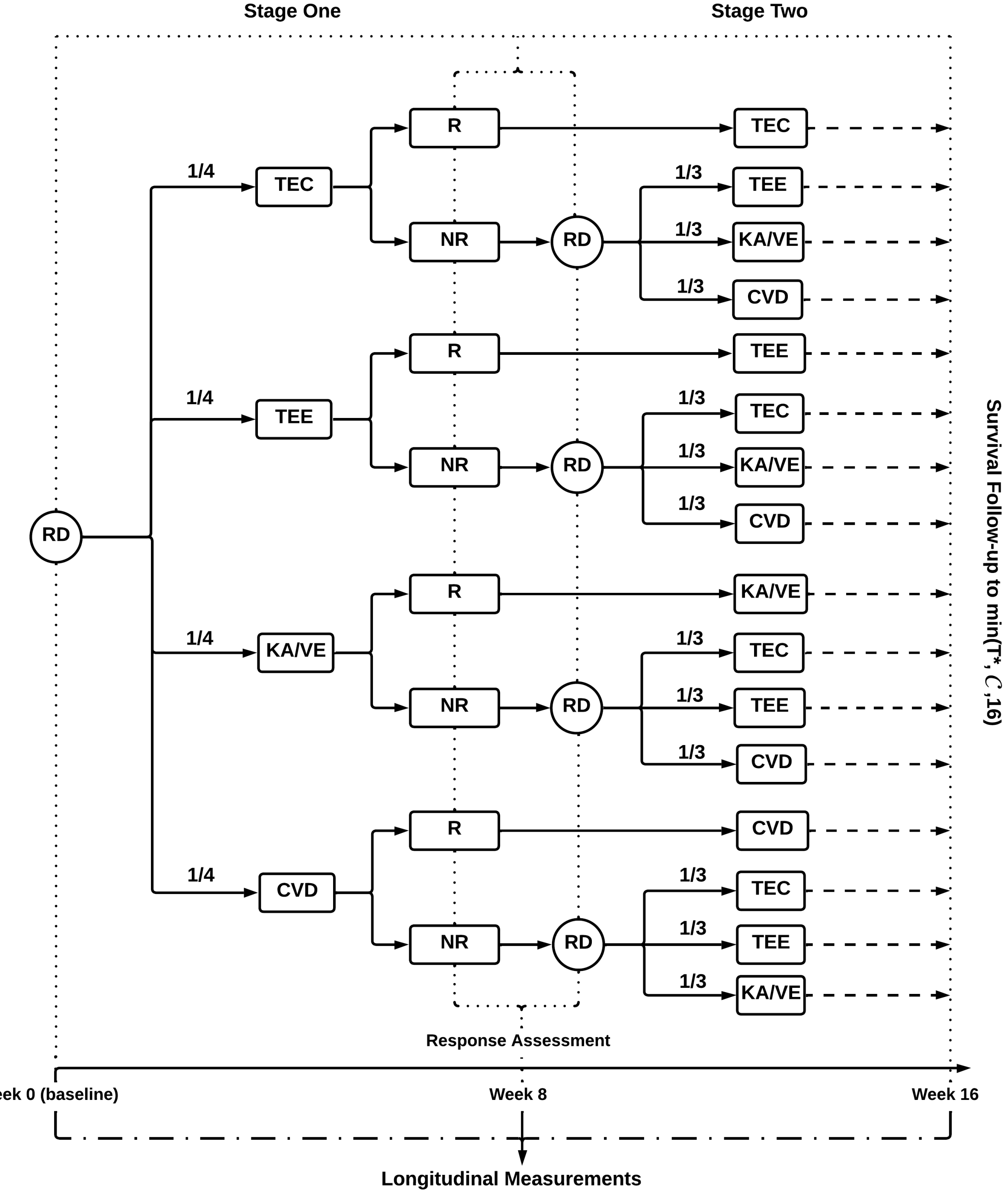

Supplementary Figure 1. Schematic of the two-stage sequential multiple assignment randomized trial design, longitudinal biomarker measurement schedule and survival follow-up for the androgen-independent prostate cancer trial. TEE, TEC, KA/VE and CVD indicate the four available treatment options; RD indicates a decision point for randomization; R and NR represent response and nonresponse; the expressions along the lines indicate treatment randomization probability; $T^*$ denotes the true event time and $C$ represents the censoring time.